\documentclass[draft,tightenlines,nofootinbib,preprint,aps,eqsecnum]{revtex4}

\newcommand{\beq}{\begin{equation}}
\newcommand{\eeq}{\end{equation}}
\newcommand{\bea}{\begin{eqnarray}}
\newcommand{\eea}{\end{eqnarray}}

\begin{document}

\title{On the Relativistic Micro-Canonical Ensemble and Relativistic Kinetic
Theory for N Relativistic Particles in Inertial and Non-Inertial Rest Frames.%
}
\author{David Alba}
\affiliation{Sezione INFN di Firenze\\
Polo Scientifico, via Sansone 1\\
50019 Sesto Fiorentino, Italy\\
E-mail alba@fi.infn.it}
\author{Horace W. Crater}
\affiliation{The University of Tennessee Space Institute \\
Tullahoma, TN 37388 USA \\
E-mail: hcrater@utsi.edu}
\author{Luca Lusanna}
\affiliation{Sezione INFN di Firenze\\
Polo Scientifico, Via Sansone 1\\
50019 Sesto Fiorentino (FI), Italy\\
E-mail: lusanna@fi.infn.it}

\begin{abstract}
A new formulation of relativistic classical mechanics allows a
reconsideration of old unsolved problems in relativistic kinetic theory and
in relativistic statistical mechanics. In particular a definition of the
relativistic micro-canonical partition function is given strictly in terms
of the Poincar\'e generators of an interacting N-particle system both in the
inertial and non-inertial rest frames. The non-relativistic limit allows a
definition of both the inertial and non-inertial micro-canonical ensemble in
terms of the Galilei generators.
\end{abstract}

\maketitle

\vfill\eject

\section{Introduction}

In Ref.\cite{1} we developed a new version of relativistic classical and
quantum mechanics (RCM and RQM) for systems of N positive-energy particles
in the rest-frame instant form of dynamics previously defined in Refs. \cite%
{2,3,4,5,6}. In it inertial frames are described by 3+1 splittings of
Minkowski space-time and the associated inertial observers use
Lorentz-scalar radar 4-coordinates $\sigma^A = (\tau, \sigma^r)$, where $%
\tau $ is the proper time of the observer atomic clock and the 3-coordinates 
$\sigma^r$ have the observer as origin. The non-relativistic limit
reproduces classical and quantum mechanics with a Hamilton-Jacobi
description of the Newtonian center of mass.

\medskip

In this formulation there is a solution to the elimination of relative times
in relativistic bound states \cite{4,5,7,8,9}, a complete control of the
relativistic collective variables and an explicit expression of the
Poincar\'e generators for isolated systems. \medskip

An isolated N-particle system with total 4-momentum $P^{\mu} = Mc\, \Big(%
\sqrt{1 + {\vec h}^2}; \vec h\Big)$, where $Mc$ is the invariant mass of the
system, is described as an \textit{external} decoupled (non-local,
non-measurable) non-covariant frozen 3-center of mass $\vec z$ ($\vec z/Mc = 
{\vec x}_{NW}(0)$ are the Cauchy data for the Newton-Wigner 3-position),
whose conjugate variable is the 3-velocity $\vec h$. This external 3-center
of mass, described by these frozen (non-evolving) Jacobi data and tending to
the Hamilton-Jacobi version of the Newton center of mass in the
non-relativistic limit, carries a \textit{pole-dipole} structure: the
invariant mass $M c$ and the rest spin $\vec S$ of the isolated system. The
Wigner 3-spaces of the \textit{inertial rest frame} are orthogonal to $%
P^{\mu}$. In them the N particles are described by Wigner spin-1 phase space
3-vectors ${\vec \eta}_i(\tau)$, ${\vec \kappa}_i(\tau)$ restricted by
rest-frame conditions eliminating the \textit{internal} center of mass.
Therefore the N-particle system is described by the external center of mass
with canonical variables $\vec z$, $\vec h$, and by N-1 relative canonical
variables ${\vec \rho}_a(\tau)$, ${\vec \pi}_a(\tau)$, $a = 1,..,N-1$. The
Lorentz-scalar invariant mass $M$ and the rest spin $\vec S$ are functions
of these relative variables and the Hamiltonian for the evolution in the
Wigner 3-spaces is $M c$.

\medskip

The non-covariant canonical 4-center of mass ${\tilde x}^{\mu}(\tau)$, the
covariant non-canonical Fokker-Pryce center of inertia $Y^{\mu}(\tau)$ and
the non-covariant non-canonical M$\o $ller center of energy $R^{\mu}(\tau)$
are the only relativistic collective variables of an isolated system which
can be built only in terms of the Poincar\'e generators of the system. They
have well defined expressions in terms of $\tau$, $\vec z$, $\vec h$, $M$, $%
\vec S$, which tend to the Newtonian center of mass for $c \rightarrow
\infty $. Only $Y^{\mu}(\tau)$ has a well defined world-line. All the
pseudo-world-lines of the other two collective variables fill a
non-covariance M$\o $ller world-tube of radius $|\vec S| / M c$ whose
properties are described in Ref.\cite{1}.

The particle world-lines $x^{\mu}_i(\tau)$ and the standard momenta $%
p^{\mu}_i(\tau)$ are derived quantities (functions of $\tau$, $\vec z$, $%
\vec h$, ${\vec \rho}_a(\tau)$, ${\vec \pi}_a(\tau)$): they turn out to be 
\textit{covariant but not canonical predictive coordinates and momenta}.

\medskip

Moreover, as shown in Ref.\cite{3}, it is possible to reformulate RCM in
global non-inertial frames of Minkowski space-time by means of \textit{%
parametrized Minkowski theories} of isolated systems (see Section II) and to
define RQM in some of them \cite{10} \footnote{%
The same happens in the \textit{parametrized Galilei theories} defined in
Ref.\cite{ 11}. This allows us to define non-relativistic quantum mechanics
(QM) in Galilean non-inertial frames.}. In this approach the transition from
a non-inertial frame to another (either non-inertial or inertial) one is
formulated as a gauge transformation. Therefore, at least at the classical
level, one has gauge equivalence of the inertial and non-inertial dynamics.
The open problem at the quantum level is the implementation of these gauge
transformations as unitary gauge transformations \footnote{%
With particle systems this can be done only in rotating systems till now 
\cite{10}. Instead the problem is completely open for the massive
Klein-Gordon field due to the no-go theorem of Ref.\cite{12} forbidding
unitary evolution in non-inertial frames.}. An important class of
non-inertial frames are the \textit{non-inertial rest frames}: in them one
can give the explicit form of all the relevant quantities defined in the
inertial rest frame.

\bigskip

In this paper we want to use the framework of this RCM for N-particle
systems, fully consistent with Lorentz signature and with the Poincar\'e
group under control, to give a definition of the relativistic
micro-canonical ensemble in relativistic statistical mechanics and of the
one-particle distribution function in relativistic kinetic theory. We want
to study the aspects of relativistic kinetic theory and relativistic
statistical mechanics (see for instance Refs. \cite{13,14}) connected with
inertial and non-inertial frames in Minkoski space-time without entering
into the foundational problems of relativistic thermodynamics, but taking
into account the problem of the relativistic center of mass and the
implications of the Wigner covariance of our 3-vectors ${\vec \eta}_i(\tau)$%
, ${\vec \kappa}_i(\tau)$, for the transformation properties of the relevant
distribution functions. Therefore we will not consider canonical and
grand-canonical ensembles, which are not equivalent to the micro-canonical
ensemble when long-range interactions are present, and we will define only
the micro-canonical temperature.

\medskip

All these topics are relevant for astrophysics, cosmology, Brownian motion,
plasma physics, heavy ion collisions and quark-gluon plasma. A very rich
bibliography on many of these arguments can be found in Ref.\cite{15}. In
Refs. \cite{3,16} there is also the inclusion of the electro-magnetic field
in the radiation gauge, but the consequences of its presence in non-inertial
frames for plasma physics (Vlasov equation) and magneto-hydrodynamics have
still to be explored. \bigskip

Our approach will clarify the following arguments:\bigskip

A) In Ref.\cite{17} (see also its bibliography) it was said that the lack of
understanding of RCM with action-at-a-distance potentials forced people to
develop a relativistic kinetic theory of "world-lines" and not of particles.
This was due to the fact that it was not known how to parametrize the
world-lines of N relativistic particles in an arbitrary inertial frame in
terms of a Lorentz scalar time-parameter so that their interactions depend
only on space-like distances between pairs of particles and not on relative
times (the basic problem in the theory of relativistic bound states). In our
new RCM all these problems are solved: it is clarified how to synchronize
the clocks of the N particles and how to formulate a Cauchy problem 
\footnote{%
Even if non physical we must give the Cauchy data on a space-like 3-space,
to be able to use the existence and unicity theorem for the solution of
partial differential equations to predict the future. The subsequent
evolution is consistent with special relativity even in the case of
action-at-a-distance interactions.} in presence of action-at-a-distance,
electro-magnetic and gravitational interactions. We will give the explicit
form of the Poincar\'e generators for a simple model of N positive-energy
particles with action-at-a-distance potentials \footnote{%
See also the 2-body model of Ref.\cite{7} and the complicated N-particle
model with Coulomb plus Darwin potentials of Refs.\cite{4,5,8,9}.}: for it
we can find the Liouville operators for single particles in inertial and
non-inertial frames. A crucial difference with respect to other approaches
is the decoupling of the external (non-local, non-measurable) relativistic
center of mass \medskip

Let us remark that in all the previous attempts to define relativistic
kinetic theory and relativistic statistical mechanics starting from a system
of N relativistic particles, like the ones in Refs.\cite{18,19,20,21} (see
also the review in Ref.\cite{22}), there are either no interactions or ad
hoc ansatzs to avoid the foundational problems of RCM (quoted in Refs.\cite%
{1,7,23}), whose validity is out of control in these approaches.

\bigskip

B) In Ref.\cite{24} there is the evaluation of the standard extended (i.e.
depending also on the conserved angular momentum and not only on the energy
of the isolated system like the ordinary one) non-relativistic distribution
function for the micro-canonical ensemble for a system of N particles
interacting with the long range Newton gravitational interaction (see Ref.%
\cite{25} for the status of long-range interactions in non-relativistic
statistical mechanics and the non-equivalence of micro-canonical and
canonical ensembles). This suggested that we look for a formulation of the
(ordinary and extended) relativistic micro-canonical ensemble both in the
inertial rest frame and in the non-inertial ones (where the inertial forces
are long range forces) for N particle systems \footnote{%
See Ref. \cite{26} for the extended micro-canonical ensemble of the ideal
relativistic \textit{quantum} gas in relativistic inertial frames.}. It
turns out that the definitions of these distribution functions depend in a
natural way on the ten internal Poincar\'e generators and that the ensemble
depends only upon the canonical relative variables ${\vec \rho}_a$, ${\vec
\pi}_a$, $a=1,..,N-1$, but not on the frozen Jacobi data of the external
center of mass. The non-relativistic limit allows us to find the (ordinary
and extended) Newtonian micro-canonical ensembles both in inertial and
non-inertial rest frames of Galilei space-time by using the generators of
the Galilei group in the Hamilton-Jacobi description of the center of mass.
Now, unlike with the relativistic case, one can reintroduce the motion of
the center of mass to recover the known definition of the distribution
function. In the relativistic case the non-covariance of the Jacobi data $%
\vec z$ makes the reintroduction of the external center of mass impossible.

We are able to evaluate explicitly our modified distribution function in the
inertial non-relativistic case, but not in the inertial relativistic case
(also in the standard approach its form is not known when $m \not= 0$ \cite%
{27}). Let us remark that to our knowledge this is the first time that one
has a definition of micro-canonical ensemble in relativistic and
non-relativistic non-inertial frames. In relativistic non-inertial frames
naive relative variables cannot be defined, but action-at-a-distance
potentials $V({\vec \rho}_a^2)$ can be described by using the Synge
world-function like in general relativity \cite{28}.

\medskip

In inertial frames the (ordinary or extended) distribution function is time
independent consistent with the standard notion of \textit{equilibrium}.
This turns out to be true also in the non-inertial rest frames due to the
fact that both the Galilei or Poincar\'e generators are asymptotic constants
of motion at spatial infinity. Therefore \textit{equilibrium} (at least in a
passive viewpoint \footnote{%
This is a different problem from how to describe equilibrium in general
relativity \cite{29}, where there are physical tidal degrees of freedom of
the gravitational field and the equivalence principle forbids the existence
of global inertial frames.}) can be defined also in non-inertial rest
frames, notwithstanding the fact that the inertial forces are long-range
independently from the type of inter-particle interactions. See also Ref.%
\cite{30} (and its bibliography) and Ref.\cite{31} for the problem of the
dependence of the constitutive relations of continuum mechanics on the
non-inertial frame in the non-relativistic framework.

\bigskip

C) By using the micro-canonical entropy it is possible to define the
micro-canonical temperature $T_{(mc)}$ (see Refs. \cite{32,33}; see Refs.%
\cite{34} for the case in which long range forces are present and $T_{(mc)}$
is the only reliable notion of temperature) and to show that it is a \textit{%
Lorentz scalar}. Therefore, when the thermodynamical limit ($N, V
\rightarrow \infty$ with $N/V = const.$, $T_{(mc)} \rightarrow T$) exists,
also the canonical temperature $T$ turns out to be a Lorentz scalar. This is
our answer to the endless debate on the transformation properties of
temperature under Lorentz boosts ($T = T_{rest}$, $T = T_{rest}\, (1 -
v^2/c^2)^{1/2}$, $T = T_{rest}\, (1 - v^2/c^2)^{-1/2}$): see Ref.\cite{35}
for the formulation of the problem and Refs.\cite{36} for recent
contributions. In non-inertial frames $T_{(mc)}$ will be a functional of the
inertial potentials. \bigskip

See Ref.\cite{37} for an extended version of this paper with many more
details, with a discussion of the obstacles to find a relativistic Boltzmann
equation in this approach and with a discussion of the state of
understanding of relativistic viscous fluids. In this paper there are also
the explicit calculations of some results quoted in this paper without
demonstration.

\bigskip

In Section II we make a review of those aspects of RCM and of its
non-relativistic limit which are needed to develop relativistic kinetic
theory and statistical mechanics. In Subsection A there is RCM in the
inertial rest-frame instant form and in Subsection B its non-relativistic
limit. In Subsection C there is RCM in non-inertial rest frames and in
Subsection D its non-relativistic limit.

\medskip

In Section III, after recalling the standard formulation of the
micro-canonical ensemble in non-relativistic inertial frames (Subsection A),
we give our new definition of it (and of its extension depending also on the
angular momentum) in the non-relativistic inertial frames (Subsection B), in
the relativistic inertial rest frames (Subsections C, D and E; in Subsection
E there is the definition of the relativistic micro-canonical temperature),
in relativistic non-inertial rest frames (Subsection F) and finally in
non-relativistic non-inertial frames (Subsection G).\medskip

Then there are the Conclusions and some discussion of the open problems.

\section{Classical Relativistic N-Body Systems in the Inertial and
Non-Inertial Rest Frames}

Let us consider an isolated system of N positive-energy scalar particles
either free or interacting with action-at-a-distance potentials.

\medskip

As shown in Ref.\cite{3} we now have a metrology-oriented description of
non-inertial frames in special relativity. This can be done with the \textit{%
3+1 point of view} and the use of observer-dependent Lorentz-scalar radar
4-coordinates. Let us give the world-line $x^{\mu}(\tau)$ of an arbitrary
time-like observer carrying a standard atomic clock: $\tau$ is an arbitrary
monotonically increasing function of the proper time of this clock. Then we
give an admissible 3+1 splitting of Minkowski space-time \footnote{%
See Ref.\cite{1} for the admissibility conditions.}, namely a nice foliation
with space-like instantaneous 3-spaces $\Sigma_{\tau}$ tending in a
direction-independent way to space-like hyper-planes (all parallel) at
spatial infinity: it is the mathematical idealization of a protocol for
clock synchronization (all the clocks in the points of $\Sigma_{\tau}$ sign
the same time of the atomic clock of the observer). On each 3-space $%
\Sigma_{\tau}$ we choose curvilinear 3-coordinates $\sigma^r$ having the
observer as origin. These are the \textit{radar 4-coordinates} $\sigma^A =
(\tau; \sigma^r)$. If $x^{\mu} \mapsto \sigma^A(x)$ is the coordinate
transformation from the Cartesian 4-coordinates $x^{\mu}$ of a reference
inertial observer to radar coordinates, its inverse $\sigma^A \mapsto
x^{\mu} = z^{\mu}(\tau ,\sigma^r)$ defines the \textit{embedding} functions $%
z^{\mu}(\tau ,\sigma^r)$ describing the 3-spaces $\Sigma_{\tau}$ as an
embedded 3-manifold into Minkowski space-time. From now on we shall denote
the curvilinear 3-coordinates $\sigma^r$ with the notation $\vec \sigma$ for
the sake of simplicity.\medskip

The induced 4-metric on $\Sigma_{\tau}$ is the following functional of the
embedding: ${}^4g_{AB}(\tau , \vec \sigma) = [z^{\mu}_A\, \eta_{\mu\nu}\,
z^{\nu}_B](\tau , \vec \sigma)$, where $z^{\mu}_A = \partial\,
z^{\mu}/\partial\, \sigma^A$ and ${}^4\eta_{\mu\nu}$ is the flat metric. The
4-metric ${}^4g_{AB}$ has signature $\epsilon\, (+---)$ with $\epsilon = \pm$
(the particle physics, $\epsilon = +$, and general relativity, $\epsilon = -$%
, conventions); the flat Minkowski metric is $\eta_{\mu\nu} = \epsilon\,
(+---)$.\medskip

While the 4-vectors $z^{\mu}_r(\tau , \vec \sigma)$ are tangent to $%
\Sigma_{\tau}$, so that the unit normal $l^{\mu}(\tau , \vec \sigma)$ is
proportional to $\epsilon^{\mu}{}_{\alpha \beta\gamma}\, [z^{\alpha}_1\,
z^{\beta}_2\, z^{\gamma}_3](\tau , \vec \sigma)$, we have $%
z^{\mu}_{\tau}(\tau , \vec \sigma) = [N\, l^{\mu} + n^r\, z^{\mu}_r](\tau ,
\vec \sigma)$ for the so-called evolution 4-vector, where $N(\tau , \vec
\sigma) = 1 + n(\tau, \vec \sigma) = \epsilon\, [z^{\mu}_{\tau}\,
l_{\mu}](\tau, \vec \sigma)$ and $n_r(\tau, \vec \sigma) = - \epsilon\,
g_{\tau r}(\tau, \vec \sigma) = [{}^3g_{rs}\, n^s](\tau, \vec \sigma)$ are
the lapse and shift functions. We also have $|det\, {}^4g| = (1 + n)\, \sqrt{%
\gamma}$; $\sqrt{\gamma} = \sqrt{det\, {}^3g}$ with ${}^3g_{rs} = -
\epsilon\, {}^4g_{rs}$ of positive signature.\medskip

\medskip

In Ref.\cite{3} there is a complete description of the isolated systems
admitting a Lagrangian description in non-inertial frames by means of 
\textit{parametrized Minkowski theories}. In them there is a well defined
action principle containing the embeddings $z^{\mu}(\tau, \vec \sigma)$ as
Lagrangian variables and allowing the determination of the energy-momentum
tensor $T^{\mu\nu}(z(\tau, \vec \sigma)) = \Big(z^{\mu}_A\, z^{\nu}_B\,
T^{AB}\Big)(\tau, \vec \sigma) = l^{\mu}\, l^{\nu}\, T_{\perp\perp} +
(l^{\mu}\, z^{\nu}_r + l^{\nu}\, z^{\mu}_r)\, h^{rs}\, T_{\perp s} +
z^{\mu}_r\, z^{\mu}_s\, T^{rs}$ ($T_{\perp\perp} = l_{\mu}\, l_{\nu}\,
T^{\mu\nu} = (1 + n)^2\, T^{\tau\tau}$, $T_{\perp r} = l_{\mu}\, z_{r\,
\nu}\, T^{\mu\nu} = - (1 + n)\, h_{rs}\, (T^{\tau\tau}\, n^s + T^{\tau s})$, 
$T_{rs} = z_{r\, \mu}\, z_{s\, \nu}\, T^{\mu\nu} = n_r\, n_s\, T^{\tau\tau}
+ (n_r\, h_{su} + n_s\, h_{ru})\, T^{\tau u} + h_{ru}\, h_{sv}\, T^{uv}$) of
the isolated system. This allows us to find the ten Poincar\'e generators
and to study the configurations of the isolated system having them finite.
We shall only consider the case in which the total 4-momentum is time-like.

\medskip

The embeddings turn out to be gauge variables (i.e. the transition from a
frame to another one is a gauge transformation), because their conjugate
canonical momenta $\rho_{\mu}(\tau, \vec \sigma)$ are determined by four
first class constraints (a de-parametrization of the super-momentum and
super-Hamiltonian constraints of general relativity; they exist due to the
invariance of the action principle under frame-preserving diffeomorphisms)

\begin{eqnarray}
\rho_{\mu}(\tau , \vec \sigma) &=& \Big(\sqrt{- g}\, z_{A\, \mu}\, T^{\tau A}%
\Big)(\tau , \vec \sigma) =  \nonumber \\
&=& \Big((1 + n)^2\, \sqrt{\gamma}\, T^{\tau\tau}\, l_{\mu} + (1 + n)\, 
\sqrt{\gamma}\, \Big[T^{\tau r} + T^{\tau\tau}\, n^r\Big]\, z_{r\, \mu}\Big)%
(\tau , \vec \sigma).  \label{2.1}
\end{eqnarray}

\medskip

The ten Poincar\'e generators are

\begin{equation}
P^{\mu} = \int d^3\sigma\, \rho^{\mu}(\tau, \vec \sigma),\qquad J^{\mu\nu} =
\int d^3\sigma\, \Big(z^{\mu}\, \rho^{\nu} - z^{\nu}\, \rho^{\mu}\Big)(\tau,
\vec \sigma).  \label{2.2}
\end{equation}

\subsection{Classical Relativistic Mechanics in the Rest-Frame Instant Form}

In special relativity we can restrict ourselves to inertial frames and
define the \textit{inertial rest-frame instant form of dynamics for isolated
systems} by choosing the 3+1 splitting corresponding to the intrinsic
inertial rest frame of the isolated system centered on an inertial observer:
the instantaneous 3-spaces, named Wigner 3-space due to the fact that the
3-vectors inside them are Wigner spin-1 3-vectors \cite{3}, are orthogonal
to the conserved 4-momentum $P^{\mu}$ of the configuration.

\bigskip

The form of the embedding of the Wigner 3-spaces in Minkowski space-time
described in the inertial frame of an arbitrary inertial observer is\medskip

\begin{eqnarray}
z_W^{\mu}(\tau, \vec \sigma) &=& Y^{\mu}(\tau) + \epsilon^{\mu}_r(\vec h)\,
\sigma^r = Y^{\mu}(0) + \Lambda^{\mu}{}_A(\vec h)\, \sigma^A,  \nonumber \\
{}&&  \nonumber \\
&&\epsilon^{\mu}_{\tau}(\vec h) = {\frac{{P^{\mu}}}{{M c}}} = u^{\mu}(P) =
h^{\mu} = \Big(\sqrt{1 + {\vec h}^2}; \vec h\Big) =
\Lambda^{\mu}{}_{\tau}(\vec h),\qquad \epsilon\, P^2 = M^2\, c^2,  \nonumber
\\
&&{}  \nonumber \\
&&\epsilon^{\mu}_r(\vec h) = \Big( h_r; \delta^i_r + {\frac{{h^i\, h_r}}{{1
+ \sqrt{1 + {\vec h}^2}}}}\Big) = \Lambda^{\mu}{}_r(\vec h),  \label{2.3}
\end{eqnarray}

\noindent where $Y^{\mu}(\tau ) = Y^{\mu}(0) + h^{\mu}\, \tau =
z^{\mu}_W(\tau ,\vec 0)$ is the world-line of the external Fokker-Pryce
4-center of inertia with $\eta_{\mu\nu}\, \epsilon^{\mu}_A(\vec h)\,
\epsilon^{\nu}_B(\vec h) = \eta_{AB}$. \bigskip

Since the 3-metric inside the Wigner 3-space is Euclidean with positive
signature we have $h^i = h_i$. The Lorentz matrix $\Lambda^{\mu}{}_A(\vec h)$
is obtained from the standard Wigner boost $\Lambda^{\mu}{}_{\nu}(P^{%
\alpha}/Mc)$, sending the time-like 4-vector $P^{\mu}/Mc$ into $(1; 0)$, by
transforming the index $\nu$ into an index adapted to radar 4-coordinates ($%
\Lambda^{\mu}{}_{\nu} \mapsto \Lambda^{\mu}{}_A$).

\bigskip

The non-canonical covariant external Fokker-Pryce 4-center of inertia
(origin of the Wigner 3-space) and the canonical non-covariant external
4-center of mass have the following expression

\begin{eqnarray}
Y^{\mu}(\tau ) &=& \Big({\tilde x}^o(\tau ); \vec Y(\tau )\Big) = \Big(\sqrt{%
1 + {\vec h}^2}\, (\tau + {\frac{{\vec h \cdot \vec z}}{{Mc}}}); {\frac{{%
\vec z}}{{Mc}}} + (\tau + {\frac{{\vec h \cdot \vec z}}{{Mc}}})\, \vec h + {%
\frac{{\vec S \times \vec h}}{{Mc\, (1 + \sqrt{1 + {\vec h}^2})}}} \Big) = 
\nonumber \\
&=& z_W^{\mu}(\tau ,\vec 0),\qquad Y^{\mu}(0) = \Big(\sqrt{1 + {\vec h}^2}\, 
{\frac{{\vec h \cdot \vec z}}{{Mc}}}; {\frac{{\vec z}}{{Mc}}} + {\frac{{\vec
h \cdot \vec z}}{{Mc}}}\, \vec h + {\frac{{\vec S \times \vec h}}{{Mc\, (1 + 
\sqrt{1 + {\vec h}^2})}}}\Big),  \nonumber \\
{}&&  \nonumber \\
{\tilde x}^{\mu}(\tau ) &=& \Big({\tilde x}^o(\tau ); {\tilde {\vec x}}(\tau
)\Big) = z^{\mu}_W(\tau ,{\tilde {\vec \sigma}}) = Y^{\mu}(\tau ) + \Big(0; {%
\frac{{- \vec S \times \vec h}}{{Mc\, (1 + \sqrt{1 + {\vec h}^2})}}}\Big),
\label{2.5}
\end{eqnarray}

\noindent with ${\tilde {\vec \sigma}} = {\frac{{- \vec S \times \vec h}}{{%
Mc\, (1 + \sqrt{1 + {\vec h}^2})}}}$ giving its location on the Wigner
3-space.\medskip

\bigskip

Every isolated system (i.e. a closed universe) can be visualized as a
decoupled non-covariant collective (non-local) pseudo-particle described by
the frozen Jacobi data $\vec z$, $\vec h$ carrying a \textit{pole-dipole
structure}, namely the invariant mass $M\, c$ and the rest spin ${\vec S}$
of the system, and with an associated \textit{external} realization of the
Poincar\'e group (the last term in the Lorentz boosts induces the Wigner
rotation of the 3-vectors inside the Wigner 3-spaces):\medskip

\begin{eqnarray}
P^{\mu} &=& M\, c\, h^{\mu} = M\, c\, \Big(\sqrt{1 + {\vec h}^2}; \vec h\Big)%
,  \nonumber \\
&&{}  \nonumber \\
J^{ij} &=& z^i\, h^j - z^j\, h^i + \epsilon^{ijk}\, S^k,\qquad K^i = J^{oi}
= - \sqrt{1 + {\vec h}^2}\, z^i + {\frac{{(\vec S \times \vec h)^i}}{{1 + 
\sqrt{1 + {\vec h}^2}}}},  \label{2.8}
\end{eqnarray}

\noindent satisfying the Poincar\'e algebra.

\medskip

Inside the Wigner 3-spaces the system is described by an internal 3-center
of mass with a conjugate 3-momentum \footnote{%
Due to the rest-frame condition ${\vec{\mathcal{P}}}\approx 0$ of Eq.(\ref%
{2.9}), we have ${\vec{q}}_{+}\approx {\vec{R}}_{+}\approx {\vec{y}}_{+}$,
where ${\vec{q}}_{+}$ is the internal canonical 3-center of mass (the
internal Newton-Wigner position), ${\vec{y}}_{+}$ is the internal
Fokker-Pryce 3-center of inertia and ${\vec{R}}_{+}$ is the internal M$\o $%
ller 3-center of energy. As a consequence there is a unique internal
3-center of mass, which is eliminated by the vanishing of the internal
Lorentz boosts.} and by relative variables and there is an \textit{%
unfaithful internal} realization of the Poincar\'{e} group (whose generators
are determined by using the energy-momentum tensor of the isolated system):
the internal 3-momentum, conjugate to the (unique -see footnote 8) internal
3-center of mass, vanishes due the rest-frame condition. To avoid a double
counting of the center of mass, i.e. an external one and an internal one,
the internal (interaction-dependent) Lorentz boosts must also vanish. The
only non-zero internal generators are the invariant mass $M\,c$ and the rest
spin ${\vec{S}}$ and the dynamics is re-expressed only in terms of internal
Wigner-covariant relative variables.

\bigskip

The generators of the unfaithful internal realization of the Poincare'
algebra determined by the energy-momentum tensor (in inertial frames Eqs.(%
\ref{2.1}) imply $T_{\perp\perp} = T^{\tau\tau}$ and $T_{\perp r} =
\delta_{rs}\, T^{\tau s}$) are

\begin{eqnarray}
Mc &=&\int d^{3}\sigma \,T^{\tau \tau }(\tau ,\vec{\sigma}),\qquad {S}^{r}={%
\frac{1}{2}}\,\delta ^{rs}\,\epsilon _{suv}\,\int d^{3}\sigma \,\sigma
^{u}\,T^{\tau v}(\tau ,\vec{\sigma}),  \nonumber \\
&&{}  \nonumber \\
\mathcal{P}^{r} &=&\int d^{3}\sigma \,T^{\tau r}(\tau ,\vec{\sigma})\approx
0,\qquad \mathcal{K}^{r}=-\int d^{3}\sigma \,\sigma ^{r}\,T^{\tau \tau
}(\tau ,\vec{\sigma})\approx 0.  \label{2.9}
\end{eqnarray}

\medskip

The constraints ${\vec {\mathcal{P}}} \approx 0$ are the \textit{rest-frame
conditions} identifying the inertial rest frame. Having chosen the
Fokker-Pryce center of inertia as origin of the 3-coordinates, the (\textit{%
interaction-dependent}) constraints ${\vec {\mathcal{K}}} \approx 0$ can be
shown \cite{1,3} to be their gauge fixing: they eliminate the internal
3-center of mass so not to have a double counting (external, internal).
Therefore the isolated system is described by the external non-covariant
3-center of mass $\vec z$, $\vec h$, and by an \textit{internal space} of
Wigner-covariant relative variables ($M$ and ${\vec {\bar S}}$ depend only
upon them).

\bigskip

For N free positive energy spinless particles their world-lines are
parametrized in terms of Wigner 3-vectors ${\vec \eta}_i(\tau)$, $i = 1,..,N$%
, in the following way ($\eta^A_i(\tau) = (\tau; \eta^r_i(\tau))$)

\begin{eqnarray}
x^{\mu}_i(\tau) &=& z^{\mu}_W(\tau, {\vec \eta}_i(\tau)) = Y^{\mu}(\tau) +
\epsilon^{\mu}_r(\vec h)\, \eta^r_i(\tau) = Y^{\mu}(0) +
\Lambda^{\mu}{}_A(\vec h)\, \eta^A_i(\tau).  \label{2.11}
\end{eqnarray}

\medskip

At the Hamiltonian level the basic canonical variables describing the
particle are ${\vec \eta}_i(\tau)$ and their canonically conjugate momenta ${%
\vec \kappa}_i(\tau)$: $\{ \eta^r_i(\tau), \kappa^s_j(\tau)\} =
\delta_{ij}\, \delta^{rs}$. The standard momenta of the positive-energy
scalar particles are

\begin{equation}
p_i^{\mu}(\tau) = \Lambda^{\mu}{}_A(\vec h)\, \kappa^A_i(\tau),\qquad
\kappa^A_i(\tau) = ( E_i(\tau); \kappa_{ir}(\tau) ).  \label{2.12}
\end{equation}

\noindent For free particles we have $E_i(\tau) = \sqrt{m_i^2\, c^2 + {\vec
\kappa}_i^2(\tau)}$, $\epsilon\, p_i^2 = m_i^2\, c^2$ and $M c = \sum_i\,
E_i $. \medskip

In the interacting case it is $E_i \not= \sqrt{m_i^2\, c^2 + {\vec \kappa}%
_i^2(\tau)}$ and $\epsilon\, p_i^2 \not= m_i^2\, c^2$. Instead $E_i(\tau)$
must be deduced from the form of the invariant mass $M c$, which is the
Hamiltonian for the $\tau$-evolution in the Wigner 3-spaces. In the two-body
model of Ref.\cite{7} and in the N-body model of Section IV of Ref.\cite{37}%
, the invariant mass has the form \footnote{%
Instead the rest spin is always $\vec S = \sum_i\, {\vec \eta}_i \times {%
\vec \kappa}_i$ being in an instant form of dynamics.} $M c =
\sum_i^{1,..,N}\, E_i(\tau)$ with $E_i(\tau) = \sqrt{m_i^2\, c^2 + {\vec
\kappa}_i^2(\tau) + V_i(\tau)}$, where $V_i(\tau) = {\tilde V}_i({\vec \eta}%
_m(\tau) - {\vec \eta}_n(\tau), {\vec \kappa}_k)$ are suitable
action-at-a-distance potentials. This implies $\epsilon\, p_i^2 = m_i^2\,
c^2 + V_i(\tau)$. In these cases we have ${\dot {\vec \eta}}_i(\tau) = \{ {%
\vec \eta}_i(\tau), M c \} = {\frac{{{\vec \kappa}_i(\tau)}}{{E_i(\tau)}}}$
and $p^{\mu}_i(\tau) = E_i(\tau)\, {\dot x}^{\mu}_i(\tau)$ with ${\dot x}%
^{\mu}_i(\tau) = h^{\mu} + \epsilon^{\mu}_r(\vec h)\, \eta^r_i(\tau)$ from
Eqs. (\ref{2.11}) and (\ref{2.5}) (so that we have $\epsilon\, {\dot x}%
^2_i(\tau) = {\frac{{m_i^2\, c^2 + V_i(\tau)}}{{E_i^2(\tau)}}}$). This
description is not in contrast with scattering theory, where $\epsilon\,
p_i^2 = m_i^2\, c^2$ holds asymptotically for the \textit{in} and \textit{out%
} free particles (no interpolating description due to Haag no-go theorem for
the interaction picture: see Ref.\cite{4} for the way out, at least at the
classical level, from this theorem in the 3+1 point of view), if the
action-at-a-distance potentials (and also interactions with electro-magnetic
fields) go to zero for large separations of the particles inside Wigner
3-spaces (the \textit{cluster separability} in action-at-a-distance
theories). In relativistic kinetic theory and in relativistic statistical
mechanics $E_i = \sqrt{m_i^2\, c^2 + {\vec \kappa}_i^2(\tau)}$ holds for a
gas of non-interacting particles, otherwise it has to be replaced with an
expression dictated by the type of the existing interactions.

\bigskip

For N free positive energy spinless particles the internal generators have
the following expression in the rest frame

\begin{eqnarray}
M\, c &=& {\frac{1}{c}}\, \mathcal{E}_{(int)} = \sum_{i=1}^N\, \sqrt{m_i^2\,
c^2 + {\vec \kappa}^2_i},\qquad {\vec {\mathcal{P}}} = \sum_{i=1}^2\, {\vec
\kappa}_i \approx 0,  \nonumber \\
\vec S &=& {\vec {\mathcal{J}}} = \sum_{i=1}^2\, {\vec \eta}_i \times {\vec
\kappa}_i,\qquad {\vec {\mathcal{K}}} = - \sum_{i=1}^2\, {\vec \eta}_i\, 
\sqrt{m_i^2\, c^2 + {\vec \kappa}_i^2} \approx 0.  \label{2.14}
\end{eqnarray}

\bigskip

In the two-body case (see Refs.\cite{3,4,5,7} for the N-body case), by
introducing the notation ${\vec \eta}_+$, ${\vec \kappa}_+ = {\vec {\mathcal{%
P}}}$, with a canonical transformation we get the following collective and
relative variables

\begin{eqnarray}
{\vec \eta}_+ &=& {\frac{{m_1}}{m}}\, {\vec \eta}_1 + {\frac{{m_2}}{m}}\, {%
\vec\eta}_2, \qquad {\vec \rho} = {\vec \eta}_1 - {\vec \eta}_2,  \nonumber
\\
{\vec \kappa}_+ &=& {\vec \kappa}_1 + {\vec \kappa}_2 \approx 0, \qquad {%
\vec \pi} = {\frac{{m_2}}{m}}\, {\vec \kappa}_1 - {\ \frac{{m_1}}{m}}\, {%
\vec \kappa}_2,  \nonumber \\
&&{}  \nonumber \\
&&{}  \nonumber \\
{\vec \eta}_i &=& {\vec \eta}_+ + (-)^{i+1}\, {\frac{{m_{i+1}}}{m}}\, {\vec
\rho}, \qquad {\vec \kappa}_i = {\frac{{m_i}}{m}}\, {\vec \kappa}_+ +
(-)^{i+1}\, {\vec \pi},  \label{2.15}
\end{eqnarray}

\noindent where we use the convention $m_3 \equiv m_1$. The collective
variable ${\vec \eta}_+(\tau )$ has to be determined in terms of ${\vec \rho}%
(\tau )$ and ${\vec \pi}(\tau )$ by means of the gauge fixings ${\vec {%
\mathcal{K}}}\, {\buildrel {def}\over =}\, - M\, {\vec R}_+ \approx 0$.
For two \textit{free} particles Eqs.(\ref{2.14}) imply (${\vec \eta}(\tau)
\approx 0$ for $m_1 = m_2$)

\begin{equation}
{\vec \eta}_+(\tau) \approx {\vec \eta}(\tau) = {\frac{{{\frac{{m_1}}{m}}\, 
\sqrt{m_2^2\, c^2 + {\vec \pi}^2(\tau)} - {\frac{{m_2}}{m}}\, \sqrt{m_1^2\,
c^2 + {\vec \pi}^2(\tau)}}}{{\sqrt{m_1^2\, c^2 + {\vec \pi}^2(\tau)} + \sqrt{%
m_2^2\, c^2 + {\vec \pi}^2(\tau)}}}}\, {\vec \rho}(\tau).  \label{2.16}
\end{equation}

\bigskip

In the interacting case the rest-frame conditions ${\vec \kappa}_+ \approx 0$
and the conditions eliminating the internal 3-center of mass ${\vec {%
\mathcal{K}}} \approx 0$ will determine ${\vec \eta}_+$ in terms of the
relative variables ${\vec \rho}$, ${\vec \pi}$ in an interaction-dependent
way. Then the relative variables satisfy Hamilton equations with the
invariant mass $M({\vec \rho}, {\vec \pi})$ as Hamiltonian and the particle
world-lines $x^{\mu}_i(\tau )$ can be rebuilt \cite{7}. See Eqs.
(2.18)-(2.20) of Ref.\cite{37} for the extension of these results to N
particles. \medskip

The position of the two positive-energy particles in each instantaneous
Wigner 3-space is identified by the intersection of the world-lines ($m_3
\equiv m_1$)

\begin{eqnarray}
x^{\mu}_i(\tau ) &\approx_{free\, case}& Y^{\mu}(\tau ) +
\epsilon^{\mu}_r(\vec h)\, {\frac{\sqrt{m_i^2\, c^2 + {\vec \pi}^2(\tau )}}{{%
\sqrt{m_1^2\, c^2 + {\vec \pi}^2(\tau )} + \sqrt{m_2^2\, c^2 + {\vec \pi}%
^2(\tau )}}}}\, \rho^r(\tau),  \nonumber \\
&&{}  \nonumber \\
&& {\vec x}_i(\tau )\, \rightarrow_{c \rightarrow \infty}\, {\vec x}%
_{(n)}(t) + (-)^{i+1}\, {\frac{{m_{i+1}}}{m}}\, {\vec r}_{(n)}(t) = {\vec x}%
_{(n)i}(t),  \nonumber \\
&&{}  \nonumber \\
&&{}  \nonumber \\
p_i^{\mu}(\tau ) &=& h^{\mu}\, \sqrt{m_i^2\, c^2 + {\vec \pi}^2(\tau )} +
(-)^{i+1}\, \epsilon^{\mu}_r(\vec h)\, \pi^r(\tau ),\qquad \epsilon\,
p^2_i(\tau ) = m_i^2\, c^2,  \label{2.17}
\end{eqnarray}

\noindent with $Y^{\mu}(\tau)$ given in Eq.(\ref{2.5}) in terms of $\vec z$, 
$\vec h$ and $\tau$. In the non-relativistic limit they identify the Newton
trajectories ${\vec x}_{(n) i}(t)$ (${\vec x}_{(n)}$ and ${\vec r}_{(n)}$
are the Newton center-of-mass and relative variable, respectively).

\medskip

In Subsection C of Section II of Ref.\cite{37} there is a detailed study of
how to re-express all these results of the inertial rest frame in the
Cartesian 4-coordinates $x^{\mu }$ of an arbitrary inertial frame. In the
3-spaces $\Sigma _{x^{o}=const.}$ each particle has a different radar proper
time, $x^{o}=x_{i}^{o}(\tau _{i})$. Only \textit{on-shell}, on the solutions
of Hamilton equations, is it possible to find $\tau _{i}=\tau _{i}(x^{o})$, $%
{\tilde{x}}_{i}^{\mu }(x^{o})=(x^{o};{\vec{\tilde{x}}}_{i}(x^{o})$, ${\tilde{%
p}}_{i}^{\mu }(x^{o})=({\tilde{p}}_{i}^{o}(x^{o});{\vec{\tilde{p}}}%
_{i}(x^{o}))$, and to study the complicated transition from the 6N+1
variables $\tau $, $\vec{z}$, $\vec{h}$, ${\vec{\rho}}_{a}(\tau )$, ${\vec{%
\pi}}_{a}(\tau )$ associated with the rest-frame Wigner 3-space $\Sigma
_{\tau }$ to the 6N+1 variables $x^{o}$, ${\vec{\tilde{x}}}_{i}(x^{o})$, ${%
\vec{\tilde{p}}}_{i}(x^{o})$, associated to the Euclidean 3-space $\Sigma
_{x^{o}}$. This explains why till now there was no satisfactory theory of
relativistic kinetic theory. All these complications derive from the spatial
non-separability induced by the decoupling of the non-covariant external
center of mass, which is an effect induced by the Lorentz signature of
Minkowski space-time. As shown in Ref.\cite{37}, only on-shell and in the
rest frame with $\vec{h}=0$ one has $x_{(cm)i}^{o}(\tau )=\tau =x_{(cm)}^{o}$
and ${\vec{x}}_{(cm)i}(\tau )={\vec{z}}_{(cm)}/Mc+{\vec{f}}_{i}({\vec{\rho}}%
_{a}(\tau ),{\vec{\pi}}_{a}(\tau )$, with an explicit dependence on the
non-covariant external center of mass ${\vec{z}}_{(cm)}$ of that frame.

\bigskip

See Section IV of Ref.\cite{37} for the Poincar\'e generators of a N-body
system with action-at-a-distance interaction and Appendix A of Ref.\cite{37}
for the case of N charged particles plus Coulomb and Darwin potentials \cite%
{4}.

\subsection{The Non-Relativistic Limit of the Inertial Rest Frame}

As shown in Ref.\cite{1} for the case N = 2, (easily extended to arbitrary
N), the non-relativistic limit of the inertial rest frame leads to the
description of N non-relativistic particles in Galilei space-time in terms
of the Newton center of mass ${\vec x}_{(n)}(t)$, with conjugate momentum ${%
\vec p}_{(n)}$, and of relative variables ${\vec \rho}_{(n)a}(t)$, ${\vec \pi%
}_{(n)}(t)$, $a = 1,..,N-1$. The \textit{non-relativistic rest frame} is
defined with the conditions ${\vec x}_{(n)}(t) \approx 0$ and ${\vec p}%
_{(n)} \approx 0$. Then one can reformulate the theory in terms of the
Newton positions ${\vec x}_{(n)i}(t)$ and momenta ${\vec p}_{(n)i}(t)$ of
the particles.

\medskip

By putting $\tau =c\,t$ and by using Eqs. (\ref{2.5}), (\ref{2.11}), one can
show that ${\tilde{x}}^{o}/c$, $Y^{o}/c$, $R^{o}/c$, $x_{i}^{o}/c$, all tend
to the absolute Newton time $t$ for $c\rightarrow \infty $. Moreover one
has: ${\vec{\tilde{x}}}(\tau ),\,\,\vec{Y}(\tau ),\,\,\vec{R}(\tau )\,{%
\rightarrow }_{c\rightarrow \infty }\,\,{\vec{x}}_{(n)}(t)$ and ${\vec{x}}%
_{NW}(0)=\vec{z}/Mc\,{\rightarrow }_{c\rightarrow \infty }\,\,{\vec{x}}%
_{(n)}(0)$. Therefore the three notions of external center of mass and all
the relativistic collective variables collapse into the Newton center of
mass. For the total momentum we have $\vec{P}={\vec{p}}_{(n)}$ and Eqs.(\ref%
{2.3}) imply $h^{\mu }\,{\rightarrow }_{c\rightarrow \infty }\,\,(1;0)$, $%
\epsilon _{r}^{\mu }(\vec{h})\,{\rightarrow }_{c\rightarrow \infty
}\,\,(0;\delta _{r}^{i})$. \medskip

The spatial part of Eq.(\ref{2.3}) becomes ${\vec{z}}_{W}(\tau ,\vec{\sigma}%
)\,{\rightarrow }_{c\rightarrow \infty }\,\,{\vec{x}}_{(n)}(t)+\vec{\sigma}$%
: the non-relativistic inertial frame is centered on the Newton center of
mass. The spatial part of the world-lines (\ref{2.11}) becomes ${\vec{x}}%
_{i}(\tau )\,{\rightarrow }_{c\rightarrow \infty }\,\,{\vec{x}}_{(n)i}(t)={%
\vec{x}}_{(n)}(t)+{\vec{\tilde{\eta}}}_{i}(t)$ ( with ${\vec{\eta}}%
_{+}\mapsto {\vec{x}}_{(n)}$), where ${\vec{\tilde{\eta}}}_{i}(t)={\vec{\eta}%
}_{i}(\tau )$ are restricted by the vanishing of the non-relativistic limit
of the internal Lorentz boosts of Eq.(\ref{2.9}), ${\vec{\mathcal{K}}}/c\,{%
\rightarrow }_{c\rightarrow \infty }\,\,-\sum_{i}\,m_{i}\,{\vec{\tilde{\eta}}%
}_{i}(t)\approx 0$, so that they define positions ${\vec{\eta}}_{(n)i}(t)$
coinciding with ${\vec{x}}_{(n)i}(t)$ in the non-relativistic rest frame.
The particle momenta ${\vec{\kappa}}_{i}(\tau )={\vec{\tilde{\kappa}}}%
_{i}(t) $ collapse into momenta ${\vec{\kappa}}_{(n)i}(t)$ restricted by the
rest-frame condition ${\vec{\mathcal{P}}}\,{\rightarrow }_{c\rightarrow
\infty }\,\,\sum_{i}\,{\vec{\kappa}}_{(n)i}(t)\approx 0$, so that they
coincide with the non-relativistic momenta ${\vec{p}}_{(n)i}(t)$ in the
non-relativistic rest frame.

\medskip

The other internal Poincar\'e generators $Mc$ and $\vec S$ (the pole-dipole
carried by the external center of mass) of Eq.(\ref{2.9}) become in the free
case ($m = \sum_i\, m_i$; ${\vec \eta}_{(n)i}(t) = {\vec x}_{(n)i}(t){|}_{{%
\vec x}_{(n)} = {\vec p}_{(n)} = 0}$; ${\vec \kappa}_{(n)i}(t) = {\vec p}%
_{(n)i}(t){|}_{{\vec x}_{(n)} = {\vec p}_{(n)} = 0}$)

\begin{eqnarray}
M\, c &{\rightarrow}_{c \rightarrow \infty}& m\, c + H_{rel},\qquad H_{rel}
= \sum_i\, {\frac{{{\vec \kappa}^2_{(n)i}(t)}}{{2 m_i}}},  \nonumber \\
&&{}  \nonumber \\
\vec S &{\rightarrow}_{c \rightarrow \infty}\,\,& \sum_i\, {\vec \eta}%
_{(n)i}(t) \times {\vec \kappa}_{(n)i}(t) = {\vec S}_{(n)}.  \label{2.36}
\end{eqnarray}

\noindent In the interacting case a potential $V({\vec \eta}_{(n)i}(t) - {%
\vec \eta}_{(n)j}(t), {\vec \kappa}_{(n)i}(t))$ will be present in the
relative Hamiltonian (or internal energy) $H_{rel}$. See Ref.\cite{37} for
the connection between the variables ${\vec x}_{(n)i}(t)$, ${\vec p}%
_{(n)i}(t)$ and ${\vec x}_{(n)}(t)$, ${\vec p}_{(n)}$, ${\vec \rho}%
_{(n)a}(t) $, ${\vec \pi}_{(n)a}(t)$. \medskip

The non-relativistic limit of the external Poincar\'e generators (\ref{2.8})
gives rise to the generators of the (external) Galilei algebra (centrally
extended with the total mass $m$)

\begin{eqnarray}
P^{o}{\rightarrow }_{c\rightarrow \infty } &&mc+E_{Galilei},\qquad
E_{Galilei}={\frac{{{\vec{p}}_{(n)}^{2}}}{{2m}}}+\sum_{i}{\frac{{{\vec{\kappa%
}}_{(n)i}(t)^{2}}}{{2m_{i}}}}=\sum_{i}\,{\frac{{{\vec{p}}_{(n)i}^{2}(t)}}{{%
2m_{i}}}},  \nonumber \\
\vec{P} &=&{\vec{p}}_{(n)}={\vec{P}}_{Galilei}  \nonumber \\
\vec{J} &=&{\vec{x}}_{(n)}(t)\times {\vec{p}}_{(n)}+{\vec{S}}%
_{(n)}=\sum_{i}\,{\vec{x}}_{(n)i}(t)\times {\vec{p}}_{(n)i}(t)={\vec{J}}%
_{Galilei},  \nonumber \\
{\frac{1}{c}}\,{\vec{K}}{\rightarrow }_{c\rightarrow \infty } &&t\,{\vec{p}}%
_{(n)}-m\,{\vec{x}}_{(n)}={\vec{K}}_{Galilei}.  \label{2.37}
\end{eqnarray}

\medskip

In the non-relativistic rest frame ${\vec p}_{(n)} \approx 0$, ${\vec x}%
_{(n)} \approx 0$, there is an unfaithful internal Galilei algebra: $E_{int}
= H_{rel} = \sum_i {\frac{{{\vec \kappa}^2_{(n)i}(t)}}{{2 m_i}}}$, $\sum_i\, 
{\vec \kappa}_{(n)i} \approx 0$, ${\vec S}_{(n)}$, ${\vec K}_{(n)} = -
\sum_i\, m_i\, {\vec \eta}_{(n)i}(t) - m\, {\vec x}_{(n)} \approx 0$.

\medskip

This implies that the generators of the external Galilei algebra (\ref{2.37}%
) in the rest frame ${\vec p}_{(n)} = 0$, with the origin in the center of
mass (${\vec x}_{(n)} = 0$), become $E_{Galilei} = E_{int} = H_{rel}$, ${%
\vec P}_{Galilei} = 0$, ${\vec J}_{Galilei} = {\vec S}_{(n)}$, ${\vec K}%
_{(n)} = 0$. But this is the form which can be obtained in this frame by
means of a canonical transformation implying the Hamilton-Jacobi description
of the center of mass ($H_{com} = {\frac{{{\vec p}^2_{(n)}}}{{2 m}}}\,
\mapsto H_{com}^{(HJ)} = 0$).\medskip

The non-relativistic limit of the relativistic N-body problem reproduces
this Hamilton-Jacobi version of the non-relativistic N-body problem \cite{1}.

\subsection{The Non-Inertial Rest Frames}

The family of \textit{non-inertial rest frames} for an isolated system
consists of all the admissible 3+1 splittings of Minkowski space-time whose
instantaneous 3-spaces $\Sigma_{\tau}$ tend to space-like hyper-planes
orthogonal to the conserved 4-momentum of the isolated system at spatial
infinity \footnote{%
See Section III of Ref.\cite{37} for a review of admissible non-inertial
frames in Minkowski space-time.}. Therefore they tend to the Wigner 3-spaces
(\ref{2.3}) of the inertial rest frame asymptotically. They are relevant
because they are the only global non-inertial frames allowed by the
equivalence principle (forbidding the existence of global inertial frames)
in canonical metric and tetrad gravity \cite{38,39}, in globally hyperbolic,
asymptotically flat (asymptotically Minkowskian) space-times without
super-translations, so to have the asymptotic ADM Poincare' group.

\bigskip

These non-inertial frames can be centered on the external Fokker-Pryce
center of inertia like the inertial ones and are described by the following
embeddings (the admissibility conditions are restrictions on the functions $%
g(\tau, \vec \sigma)$ and $g^r(\tau, \vec \sigma)$)

\begin{eqnarray}
&&z^{\mu}(\tau , \vec \sigma )\approx z^{\mu}_F(\tau , \vec \sigma ) =
Y^{\mu}(\tau ) + u^{\mu}(\vec h)\, g(\tau , \vec \sigma ) +
\epsilon^{\mu}_r(\vec h)\, [\sigma^r + g^r(\tau , \vec \sigma )],  \nonumber
\\
&&  \nonumber \\
&&\qquad{\rightarrow_{|\vec \sigma |\, \rightarrow \infty}} z^{\mu}_W(\tau ,
\vec \sigma ) = Y^{\mu}(\tau ) + \epsilon^{\mu}_r(\vec h)\, \sigma^r,\qquad
x^{\mu}(\tau ) = z^{\mu}_F(\tau , 0),  \nonumber \\
&&{}  \nonumber \\
&&g(\tau ,0) = g^r(\tau ,0) = 0,\qquad g(\tau , \vec \sigma )\,
\rightarrow_{|\vec \sigma | \rightarrow \infty}\, 0,\qquad g^r(\tau , \vec
\sigma )\, \rightarrow_{|\vec \sigma | \rightarrow \infty}\, 0.  \label{3.7}
\end{eqnarray}

\bigskip

As shown in Ref.\cite{3} the internal mass and the rest-frame conditions
become (Eqs.(\ref{2.9}) are recovered for the inertial rest frame; $%
T_F^{\mu\nu} = T^{\mu\nu}|_{z_F^{1rho}}$)

\begin{eqnarray}
Mc &=& \int d^3\sigma\, \Big({\frac{{det\, (\delta^s_r + \partial_r\, g^s)}}{%
\sqrt{\gamma}}}\, T_{F\, \perp\perp} - \partial_r\, g\, h_F^{rs}\, T_{F\,
\perp s}\Big)(\tau , \vec \sigma ),  \nonumber \\
&&{}  \nonumber \\
{\hat {\mathcal{P}}}^u &=& \int d^3\sigma\, \Big( - {\frac{{\delta^{ua}\,
\epsilon_{asr}\, \epsilon_{vwt}\, \partial_v\, g\,\, \partial_w\, g^s\,\,
\partial_t\, g^r} }{\sqrt{\gamma_F}}}\, T_{F\, \perp\perp} -  \nonumber \\
&-& (\delta^u_r + \partial_r\, g^u)\, h_F^{rs}\, T_{F\, \perp s}\Big)(\tau ,
\vec \sigma ) \approx 0.  \nonumber \\
&&{}  \label{3.9}
\end{eqnarray}

\bigskip

Then it can be shown \cite{3} that the second of Eqs.(\ref{2.2}) together
with Eqs. (\ref{2.8}) and (\ref{2.9}) imply the following form of the
constraints eliminating the 3-center of mass and of the effective spin
tending to the rest spin $\vec{S}$ in the inertial rest frame

\begin{eqnarray}
{\hat {\mathcal{K}}}^u &=& \int d^3\sigma\, \Big(g\, \Big[\delta^{ur}\,
\partial_r\, g\, T_{F\, \perp\perp} - (\delta^u_r + \partial_r\, g^u)\,
h_F^{rs}\, T_{F\, \perp s}\Big] -  \nonumber \\
&-& (\sigma^u + g^u)\, \Big[{\frac{{det\, (\delta^s_r + \partial_r\, g^s)}}{%
\sqrt{\gamma}}}\, T_{F\, \perp\perp} - \partial_r\, g\, h_F^{rs}\, T_{F\,
\perp s}\Big]\Big)(\tau , \vec \sigma ) \approx 0,  \nonumber \\
&&{}  \nonumber \\
{\tilde S}^r &\approx& {\hat S}^r = {\frac{1}{2}}\, \delta^{rn}\,
\epsilon_{nuv}\, \int d^3\sigma\, \Big((\sigma^u + g^u)\, \Big[\delta^{vm}\,
\partial_m\, g\, T_{F\, \perp\perp} - (\delta^v_r + \partial_r\, g^v)\,
h_F^{rs}\, T_{F\, \perp s}\Big] -  \nonumber \\
&-&(\sigma^v + g^v)\, \Big[\delta^{um}\, \partial_m\, g\, T_{F\, \perp\perp}
- (\delta^u_r + \partial_r\, g^u)\, h_F^{rs}\, T_{F\, \perp s}\Big] \Big)%
(\tau , \vec \sigma ).  \label{3.10}
\end{eqnarray}

\noindent and these formulas allow one to recover Eqs.(\ref{2.9}) of the
inertial rest frame. \medskip

Therefore the non-inertial rest-frame instant form of dynamics is well
defined since we have: a) a decoupled center of mass carrying a pole-dipole
structure; b) well defined internal Poincar\'e generators $Mc$, ${\vec {%
\mathcal{P}}} \approx 0$, ${\vec {\tilde S}}$, ${\vec {\mathcal{K}}} \approx
0$ at spatial infinity; c) non-Euclidean 3-spaces tending in a
direction-independent way to space-like hyper-planes, where they are
orthogonal to $P^{\mu}$. In Ref.\cite{3} there is the determination of the
effective Hamiltonian $\mathcal{M}\, c$ of the non-inertial rest-frame
instant form, which is not the internal mass $Mc$, since $Mc$ describes the
evolution from the point of view of the asymptotic inertial observers. There
is an additional term interpretable as an inertial potential producing
relativistic inertial effects and the final form of the effective
Hamiltonian is

\begin{eqnarray}
\mathcal{M}\, c&=& \int d^3\sigma\, \sqrt{\gamma (\tau, \sigma^u)}\, \Big((1
+ n_F)\, T_{F\, \perp\perp} + n_F^r\, T_{F\, \perp\,r}\Big)(\tau , \vec
\sigma ).  \label{3.11}
\end{eqnarray}

\subsection{The Non-Relativistic Limit of the Non-Inertial Rest Frame}

In Ref.\cite{11} there is the non-relativistic version of parametrized
Minkowski theories. If $x^a$ are the Cartesian coordinates of an inertial
frame in Galilei space-time centered on an inertial observer, a global
non-inertial frame can be defined by means of an invertible, global
coordinate transformation ($t$ is the absolute Newton time; $a = 1,2,3$; $%
\mathcal{A}^a(t, \vec \sigma)\, \rightarrow_{|\vec \sigma| \rightarrow
\infty}\, A^a{}_b(t)\, \sigma^b$)

\begin{equation}
x^a = \mathcal{A}^a(t, \vec \sigma),\quad with\,\, inverse\quad \sigma^r = 
\mathcal{S}^r(t, \vec x).  \label{3.13}
\end{equation}

\noindent The Jacobian of this transformation and its inverse are

\begin{eqnarray}
J^a{}_r(t, \vec \sigma) &=& {\frac{{\partial\, \mathcal{A}^a(t, \vec \sigma)}
}{{\partial\, \sigma^r}}},\qquad det\, J(t,\vec \sigma) > 0,  \nonumber \\
{\tilde J}^r{}_a(t, \vec \sigma) &=& \Big[{\frac{{\partial\, \mathcal{S}%
^r(t, \vec x}}{{\partial\, x^a}}}\Big]_{\vec x = {\vec {\mathcal{S}}}(t,
\vec \sigma)},  \nonumber \\
&& J^a{}_r(t,\vec \sigma)\, {\tilde J}^r{}_b(t, \vec \sigma) =
\delta^a_b,\qquad {\tilde J}^s{}_a(t, \vec \sigma)\, J^a{}_r(t, \vec \sigma)
= \delta^s_r.  \label{3.15}
\end{eqnarray}

The non-inertial frame is centered on an accelerated observer ${\vec x}_o(t)
= {\vec {\mathcal{A}}}(t, 0)$.

\medskip

In Ref.\cite{11} there is the definition of \textit{parametrized Galilei
theories} for isolated particle systems. The Lagrangian depends on the
particle positions ${\vec {\tilde \eta}}_i(t)$ and on the functions ${\vec {%
\mathcal{A}}}(t, \vec \sigma)$ as Lagrangian variables. Since the action is
invariant under 3-diffeomorphisms the momenta $\vec \rho(t, \vec \sigma)$,
conjugate to the variables ${\vec {\mathcal{A}}}(t, \vec \sigma)$, are
determined by three first-class constraints (like Eqs.(\ref{2.1}) of the
relativistic case): $\rho^a(t, \vec \sigma) \approx \sum_i\,
\delta^3(\sigma^r - {\tilde \eta}^r_i(t))\, {\tilde J}^r{}_a(t, \vec
\sigma)\, p_{ir}(t)$, where ${\vec p}_i(t)$ are the particle momenta.
Therefore, the variables ${\vec {\mathcal{A}}}(t, \vec \sigma)$ are gauge
variables: a change of frame is a gauge transformation.

\medskip

The resulting Galilei generators are

\begin{eqnarray}
E_{Galilei} &=& H_c = \sum_{iars}\, {\frac{1}{{2 m_i}}}\, {\tilde J}%
^r{}_a(t, {\vec {\tilde \eta}}_i(t))\, p_{ir}(t)\, {\tilde J}^s{}_a(t, {%
\vec {\tilde \eta}}_i(t))\, p_{is}(t),  \nonumber \\
P^a_{Galilei} &=& \sum_{ir}\, {\tilde J}^r{}_a(t, {\vec {\tilde \eta}}%
_i(t))\, p_{ir}(t),  \nonumber \\
J^a_{Galilei} &=& {\frac{1}{2}}\, \sum_{bcd}\, \epsilon^{abc}\, \sum_i\, %
\Big[\mathcal{A}^b(t, {\vec {\tilde \eta}}_i(t))\, \delta^{cd} - \mathcal{A}%
^c(t, {\vec {\tilde \eta}}_i(t))\, \delta^{bd}\Big]  \nonumber \\
&&{\tilde J}^r{}_d(t, {\vec {\tilde \eta}}_i(t))\, p_{ir}(t),  \nonumber \\
K^a_{Galilei} &=& - \sum_i\, m_i\, \mathcal{A}^a(t, {\vec {\tilde \eta}}%
_i(t)),  \label{3.16}
\end{eqnarray}

\noindent where $H_c$ is the canonical Hamiltonian. Instead the effective
non-inertial Hamiltonian is

\begin{equation}
\mathcal{M}_{Galilei} = E_{Galilei} - \sum_i\, J^r{}_a(t, {\vec {\tilde \eta}%
}_i(t))\, {\frac{{\partial\, \mathcal{A}^a(t, {\vec {\tilde \eta}}_i(t))}}{{%
\partial\, t}}}{|}_{{\vec {\tilde \eta}}_i(t)}\,\, p_{ir}(t).  \label{3.17}
\end{equation}

\noindent When one uses the functions $\mathcal{A}^{a}(t,\vec{\sigma}%
)=x_{o}^{a}(t)+\sigma ^{r}\,R_{ra}(t)$, one recovers the standard Euler,
Jacobi, Coriolis and centrifugal forces in the equation of motion of the
particles (see Eq.(4.9) of Ref.\cite{11}). One can check that the Galilei
generators (\ref{3.16}) are constants of the motion.

\bigskip

Let us now consider the non-relativistic limit of the non-inertial rest
frame, whose embedding is given in Eq.(\ref{3.7}). We put $\tau = c\, t$ and 
${\vec \eta}_i(\tau) = {\vec {\tilde \eta}}_i(t) = {\vec x}_{(n)i}(t){|}_{{%
\vec x}_{(n)}(t) = {\vec p}_{(n)} = 0}$, ${\vec \kappa}_i(\tau) = {\vec {%
\tilde \kappa}}_i(t) = {\vec p}_{(n)i}(t){|}_{{\vec x}_{(n)}(t) = {\vec p}%
_{(n)} = 0}$. \medskip

The non-relativistic limit of the embedding (\ref{3.7}) can be done by
putting $\vec h = {\frac{{\vec v}}{c}} + O(c^{-2})$ and $\sigma^r +
g^r(\tau, \vec \sigma) \, {\rightarrow}_{c \rightarrow \infty}\,\, \mathcal{A%
}^r(t, \vec \sigma)$ and by assuming $g(\tau, \vec \sigma) = O(c^{-2})$.
Then we get

\begin{eqnarray}
{\frac{1}{c}}\, z^o(\tau, \vec \sigma) &{\rightarrow}_{c \rightarrow
\infty}& t,  \nonumber \\
\vec z(\tau, \vec \sigma) &{\rightarrow}_{c \rightarrow \infty}& {\vec y}_o
+ \vec v\, t + {\vec {\mathcal{A}}}(t, \vec \sigma).  \label{3.18}
\end{eqnarray}

If in Eq.(\ref{3.13}) we put ${\vec x}_o(t) = {\vec {\mathcal{A}}}(t,\vec{0}%
) = 0$, we see that we are in a non-inertial frame centered on the Newton
center of mass ${\vec x}_{(n)}(t) = {\vec y}_o + \vec v\, t$; if we put ${%
\vec p}_{(n)} = 0$ we are in a non-relativistic non-inertial rest
frame.\medskip

The previous conditions imply the following expression for the induced
3-metric on the 3-space

\begin{eqnarray}
h_{rs}(\tau, \vec \sigma) &=& - \epsilon\, g_{rs}(\tau, \vec \sigma) =
H_{rs}(t, \vec \sigma) + O(c^{-2}),  \nonumber \\
&& H_{rs} = \delta_{rs} + {\frac{{\partial g^r}}{{\partial\, \sigma^s}}} + {%
\frac{{\partial g^s}}{{\partial\, \sigma^r}}} + \sum_u\, {\frac{{\partial g^u%
}}{{\partial\, \sigma^r}}}\, {\frac{{\partial g^u}}{{\partial\, \sigma^s}}},
\nonumber \\
&&\sqrt{\gamma} = \sqrt{H_{rs}} + O(c^{-2}).  \label{3.19}
\end{eqnarray}

Then Eqs.(\ref{3.15}) and (\ref{3.19}) imply ($H^{rs}$ is the inverse of $%
H_{rs}$)

\begin{eqnarray}
J^a{}_r &=& \delta^a_r + {\frac{{\partial g^a}}{{\partial\, \sigma^r}}}%
,\quad with\, inverse\, {\tilde J}^r{}_a,  \nonumber \\
&&{}  \nonumber \\
H_{rs} &=& \sum_a\, J^a{}_r\, J^a{}_s,\qquad H^{rs} = \sum_a\, {\tilde J}%
^r{}_a\, {\tilde J}^s{}_a.  \label{3.20}
\end{eqnarray}

\medskip

With these notations the non-relativistic limit of the external and internal
Poincar\'e generators (\ref{2.8}), (\ref{3.9}), (\ref{3.10}) produces the
following form for the Galilei generators (\ref{3.16})

\begin{eqnarray*}
P^o &{\rightarrow}_{c \rightarrow \infty}& m c + E_{Galilei}, \qquad
E_{Galilei} = H_c = {\frac{{{\vec p}_{(n)}^2}}{{2 m}}} + \mathcal{E}%
_{Galilei},  \nonumber \\
\vec P &=& {\vec p}_{(n)} = {\vec P}_{Galilei},  \nonumber \\
\vec J &=& {\vec x}_{(n)}(t) \times {\vec p}_{(n)} + {\vec {\mathcal{S}}}%
_{Galilei} = {\vec J}_{Galilei},  \nonumber \\
{\frac{1}{c}}\, {\vec K} &{\rightarrow}_{c \rightarrow \infty}& t\, {\vec p}%
_{(n)} - m\, {\vec x}_{(n)} = {\vec K}_{Galilei},
\end{eqnarray*}

\begin{eqnarray}
\mathcal{E}_{Galilei} &=&\sum_{i}\,{\frac{1}{{2m_{i}}}}H^{rs}(t,{\vec{\tilde{%
\eta}}}_{i}(t))\,{\tilde{\kappa}}_{ir}(t)\,{\tilde{\kappa}}_{is}(t), 
\nonumber \\
\mathcal{S}_{Galilei}^{r} &=&\sum_{uv}\,\epsilon ^{ruv}\,\sum_{i}\,\mathcal{A%
}^{u}(t,{\vec{\tilde{\eta}}}_{i}(t))\,J^{v}{}_{r}(t,{\vec{\tilde{\eta}}}%
_{i}(t))  \nonumber \\
&&{\times }\sum_{s}\,H^{rs}(t,{\vec{\tilde{\eta}}}_{i}(t))\,{\tilde{\kappa}}%
_{is}(t),  \nonumber \\
&&{}  \nonumber \\
\mathcal{P}_{Galilei}^{r} &=&\sum_{ius}\,J^{r}{}_{u}(t,{\vec{\tilde{\eta}}}%
_{i}(t))\,H^{rs}(t,{\vec{\tilde{\eta}}}_{i}(t))\,{\tilde{\kappa}}%
_{is}(t)\approx 0,  \nonumber \\
\mathcal{K}_{Galilei}^{r} &=&-\sum_{i}\,m_{i}\,\mathcal{A}^{r}(t,{\vec{%
\tilde{\eta}}}_{i}(t))\approx 0.  \label{3.21}
\end{eqnarray}

\medskip

The time constancy of the generators (\ref{3.16}) implies the time constancy
of the internal Galilei generators $\mathcal{E}_{Galilei}$, ${\vec {\mathcal{%
S}}}_{Galilei}$, ${\vec {\mathcal{P}}}_{Galilei}$, ${\vec {\mathcal{K}}}%
_{Galilei}$.

\section{The Micro-Canonical Ensemble}

In this Section we will give a definition of the micro-canonical ensemble
for an isolated system of $N$ particles with arbitrary either short- or
long-range interactions both in non-relativistic and relativistic classical
mechanics and both in inertial and non-inertial rest frames. In this way we
can get a description of the equilibrium configurations of the system both
in non-relativistic and relativistic statistical mechanics not only in
inertial rest frames but also in the non-inertial ones.\medskip

Firstly in Subsection A we recall the standard definition of the
non-relativistic ordinary micro-canonical partition function $Z_{(nr,st)}(E,
V, N)$ (see for instance ch.6 of Ref.\cite{13}) and of its extended form ${%
\tilde Z}_{(nr,st)}(E, {\vec {\mathcal{S}}}, V, N)$ used in Ref.\cite{24} ($%
Z_{(nr,st)}(E, V, N) = \int d^3\mathcal{S}\, {\tilde Z}_{(nr,st)}(E, {\vec {%
\mathcal{S}}}, V, N)$), depending not only on the volume $V$, on the
particle number $N$ and on the value $E$ of the total conserved energy ($H =
E$, where $H$ is the non-relativistic Hamiltonian of the isolated system)
but also on the value ${\vec {\mathcal{S}}}$ of the total conserved angular
momentum $\vec J$. \medskip

Then in Subsection B both the ordinary and extended micro-canonical
partition functions, ${\tilde Z}_{(nr)}(\mathcal{E}, V, N)$ and ${\tilde Z}%
_{(nr)}(\mathcal{E}, {\vec {\mathcal{S}}}, V, N)$, are defined in the 
\textit{non-relativistic inertial rest frame} defined in Subsection B of
Section II, where the center of mass of the isolated system is put at the
origin of the 3-coordinates (${\vec x}_{(nr)} = 0$) and the total energy $E
= {\frac{{{\vec p}_{(nr)}^2}}{{2\, N\, m}}} + \mathcal{E}$ is replaced with
the internal energy $\mathcal{E}$, which is an invariant of the centrally
extended Galilei algebra, due to the rest-frame condition ${\vec p}_{(nr)}
\approx 0$. These definitions use only the internal Galilei generators of
the isolated system. It is possible to reintroduce the dependence on the
center of mass and to recover the standard partition functions. \medskip

In Subsection C there is the main new result, namely the definition of the
ordinary and extended micro-canonical partition functions, $\tilde Z(%
\mathcal{E}, V, N)$ and ${\tilde Z}(\mathcal{E}, {\vec {\mathcal{S}}}, V, N)$%
, in the \textit{relativistic rest frame} of Section II. In Subsection D we
study their transformation properties under Lorentz transformations. These
definitions use only the internal Poincar\'e generators of the isolated
system and do not depend on the decoupled external relativistic center of
mass. Now $\mathcal{E}$ is the conserved invariant mass $M c$ of the
isolated system (${\vec {\mathcal{S}}}$ is its rest spin) and it is not
possible to reintroduce a dependence on the external center of mass as was
possible in the non-relativistic case due to the non-covariance of the
Jacobi data $\vec z$.\medskip

Both in the non-relativistic and relativistic cases we define the \textit{%
micro-canonical temperature} (see Subsection E) and we show that it is a 
\textit{Lorentz-scalar}. We do not discuss the canonical ensemble and the
canonical temperature because we consider both short- and long-range
interactions among the particles so that in general these ensembles are not
equivalent to the micro-canonical ones. \medskip

Finally in Subsection F the two micro-canonical partition functions are
defined in the \textit{relativistic non-inertial rest frame} of Subsection C
of Section II and it is shown how to make the non-relativistic limit to the 
\textit{non-relativistic non-inertial rest frame} in Subsection G. These
non-inertial partitions functions do not seem to have been defined till now.
We introduce the problem of the notion of \textit{non-inertial equilibrium}
and of its gauge equivalence to \textit{inertial equilibrium} at least in
the passive viewpoint. \medskip

We give only the results of some long calculations displayed in Appendix C
of Ref. \cite{37}.

\subsection{The Standard Micro-Canonical Ensemble in Non-Relativistic
Inertial Frames}

The standard non-relativistic micro-canonical distribution function $f_{(mc,
nr, st)}({\vec x}_1,.., {\vec p}_N | E, V, N)$ of a system of N particles
with Hamiltonian $H$ is defined in a spatial volume $V$ \footnote{%
There two points of view regarding the volume: a) it is non-dynamical (one
considers only the motions of the isolated system of particles contained in
it); b) it is dynamical (the particles have elastic reflections at the
boundaries of the volume, so that they are not an isolated system). We
consider only the case of isolated systems as it is done in Ref.\cite{24}.}
in the following way

\begin{eqnarray}
f_{(mc, nr, st)}({\vec x}_1,.., {\vec p}_N | E, V, N) &=& {\frac{1}{{Z_{(nr,
st)}(E, V, N)}}}\, \chi(V)\, \delta (H_N({\vec x}_1, .., {\vec p}_N) - E), 
\nonumber \\
&&{}  \nonumber \\
&&\chi(V) = 1\quad for\,\, {\vec x}_i \in V,\qquad \chi(V) = 0\quad for\,\, {%
\vec x}_i \notin V,  \nonumber \\
&&{}  \nonumber \\
Z_{(nr,st)}(E, V, N) &=& \int d\Gamma_N\, \chi(V)\, \delta (H_N({\vec x}%
_1,.., {\vec p}_N) - E) = {\frac{{\partial\, \Omega_{(nr, st)}(E, V, N)}}{{%
\partial E}}},  \nonumber \\
&&\qquad \Omega_{(nr, st)}(E, V, N) = \int d\Gamma_N\, \chi(v)\, \theta (H({%
\vec x}_1,.., {\vec p}_N) - E),  \nonumber \\
&&\qquad d\Gamma_N = (N!)^{-1}\, \prod_{i=1,..,N}\, d^3x_i\, d^3p_i, 
\nonumber \\
&&\int d\Gamma_n\, f_{(mc, nr, st)}({\vec x}_1,.., {\vec p}_N| E, V, N) = 1.
\label{5.1}
\end{eqnarray}

\noindent $Z_{(nr,st)}(E, V, N)$ is the standard micro-canonical partition
function.

\medskip

Due to the Hamilton equations of the particles, it satisfies ${\frac{{%
\partial\, f_{(mc, nr, st)}}}{{\partial\, t}}} + \{ f_{(mc, nr, st)}, H \} = 
{\hat L}\, f_{(mc, nr, st)} = 0$ (Liouville theorem), where $\hat L = {\frac{%
{\partial}}{{\partial\, t}}} + \sum_i\, \Big({\frac{{\partial\, H}}{{%
\partial\, {\vec p}_i}}} \cdot {\frac{{\partial}}{{\partial\, {\vec x}_i}}}
- {\frac{{\partial\, H}}{{\partial\, {\vec x}_i}}} \cdot {\frac{{\partial}}{{%
\partial\, {\vec p}_i}}}\Big)$ is the Liouville operator. Since the system
is isolated we have ${\frac{{\partial\, f_{(mc, nr, st)}}}{{\partial\, t}}}
= 0$, so that we are in the framework of \textit{equilibrium} statistical
mechanics.

\medskip

The statistical average of a function $F(\vec{x},\vec{p};{\vec{x}}_{i},{\vec{%
p}}_{i})$ in the micro-canonical ensemble depends on an extra test particle $%
\vec{x},\vec{p}$ whose statistical behavior is to be explored

\begin{equation}
F_{(mc)}(\vec x, \vec p | E, V, N) = < F >_{(mc)} = \int d\Gamma_N\, F(\vec
x, \vec p; {\vec x}_i, {\vec p}_i)\, f_{(mc, nr, st)}({\vec x}_1,.., {\vec p}%
_N | E, V, N).  \label{5.2}
\end{equation}

As shown in Eq.(C5) of Appendix C of Ref.\cite{37} (compare with Ref.\cite%
{40}) for N free particles of mass $m$ ($H=\sum_{i=1}^{\infty }\,{\frac{{{%
\vec{p}}_{i}^{2}}}{{2m}}}$) we have the following expression for the
micro-canonical distribution function (with a spherical volume $V={\frac{4}{3%
}}\,\pi \,R^{3}$: $\chi (V)=0$ for ${\vec{x}}_{i}^{2}>R^{2}$ and $\chi (V)=1$
for ${\vec{x}}_{i}^{2}<R^{2}$)

\begin{equation}
Z_{(nr, st)}(E, V, N) = \frac{1}{N!}\frac{\left( \sqrt{2\pi m}\right)^{3N}\,
E^{3N/2}\, V^N }{E \Gamma (3N/2)}\, \theta (E).  \label{5.3}
\end{equation}

\bigskip

The standard extended partition function ${\tilde Z}_{(nr, st)}(E, {\vec {%
\mathcal{S}}}, V, N)$ is introduced in Ref.\cite{24} for the case of a
Hamiltonian with long range Newtonian gravity interactions (see Eq.(B1) of
Appendix B of Ref.\cite{37}; in this Appendix there is also the
Post-Minkowskian extension of these results \cite{39} in general relativity)
without going to the rest frame. It is defined starting from the following
extended distribution function\medskip

\begin{equation}
{\tilde f}_{(mc, nr, st)}({\vec x}_1,..., {\vec p}_N| E, {\vec {\mathcal{S}}}%
, V, N) = {\tilde Z}_{(nr, st)}^{-1}(E, {\vec {\mathcal{S}}}, V, N)\,
\chi(V)\, \delta (H_N({\vec x}_1,.., {\vec p}_N) - E)\, \delta^3({\vec S}_N
- {\vec {\mathcal{S}}}),  \label{5.4}
\end{equation}

\noindent with $Z_{(nr, st)}(E, {\vec {\mathcal{S}}}, V, N) = \int
d\Gamma_N\, \chi(V)\, \delta (H_N({\vec x}_1,.., {\vec p}_N) - E)\, \delta^3(%
{\vec S}_N - {\vec {\mathcal{S}}})$ (see Eq.(\ref{5.10}) for its expression).

\subsection{The Micro-Canonical Ensemble in the Non-Relativistic Inertial
Rest Frame}

Let us now reformulate the micro-canonical ordinary and extended partition
functions in the non-relativistic rest frame of the N-particle system, in
which the Newtonian center of mass is at rest and is chosen as the origin of
the 3-coordinates of the Euclidean 3-spaces by using the generators of the
Galilei group given in Eqs.(\ref{2.37}). The rest frame conditions are ${%
\vec P}_{Galilei} = 0$ and ${\vec K}_{Galilei} = 0$ (implying ${\vec x}%
_{(n)} = {\vec p}_{(n)} = 0$). \medskip

From Eqs (\ref{2.36}) and (\ref{2.37}) we get the following expression for
the extended and ordinary micro-canonical partition functions for an
isolated system of N particles($- {\frac{1}{m}}\, {\vec {\mathcal{K}}}%
_{Galilei, N}{|}_{{\vec {\mathcal{P}}}_{Galilei, N} = 0}\, = {\vec x}_{(n)}
= 0$; the volume $V$ is assumed to have the center in the origin; ${\vec \rho%
}_{(n)a}$ and ${\vec \pi}_{(n)a}$, $a=1,..,N-1$ are relative variables)

\begin{eqnarray*}
{\tilde Z}_{(nr)}(\mathcal{E}, {\vec {\mathcal{S}}}, V, N) &=& {\frac{1}{{N!}%
}}\, \int\, \prod^{1..N}_i\, d^3\eta_i\, \chi(V)\, \int\, \prod^{1..N}_j\,
d^3\kappa_j\, \delta(E_{Galilei, N} - \mathcal{E})  \nonumber \\
&&\delta^3({\vec S}_{Galilei, N} - {\vec {\mathcal{S}}})\, \delta^3({\vec {%
\mathcal{P}}}_{Galilei, N})\, \delta^3({\frac{{{\vec {\mathcal{K}}}%
_{Galilei, N}}}{{m}}}) =  \nonumber \\
&=&{\frac{1}{{N!}}}\, \int\, \prod^{1..N-1}_a\, d^3\rho_{(n)a}\, \chi(V)\,
\int \prod_b^{1..N-1}\, d^3\pi_{(n)b}\,  \nonumber \\
&&\delta(E_{Galilei, N} - \mathcal{E})\, \delta^3({\vec S}_{Galilei, N} - {%
\vec {\mathcal{S}}}),
\end{eqnarray*}

\begin{eqnarray}
{\tilde Z}_{(nr)}(\mathcal{E}, V, N) &=& \int d^3\mathcal{S}\, {\tilde Z}%
_{(nr)}(\mathcal{E}, {\vec {\mathcal{S}}}, V, N) =  \nonumber \\
&=& {\frac{1}{{N!}}}\, \int\, \prod^{1..N}_i\, d^3\eta_i\, \chi(V)\, \int\,
\prod^{1..N}_j\, d^3\kappa_j\, \delta(E_{Galilei, N} - \mathcal{E}) 
\nonumber \\
&& \delta^3({\vec {\mathcal{P}}}_{Galilei, N})\, \delta^3({\frac{{{\vec {%
\mathcal{K}}}_{Galilei, N}}}{{m}}}) =  \nonumber \\
&=&{\frac{1}{{N!}}}\, \int\, \prod^{1..N-1}_a\, d^3\rho_{(n)a}\, \chi(V)\,
\int \prod_b^{1..N-1}\, d^3\pi_{(n)b}\,  \nonumber \\
&&\delta(E_{Galilei, N} - \mathcal{E})  \label{5.5}
\end{eqnarray}

\medskip

Since at the non-relativistic level it is possible to find a canonical basis
of relative variables such that $E_{Galilei, N} = {\frac{1}{{2m}}}\,
\sum_a^{1..N-1}\, {\vec \pi}_a^2 +\, potentials$, the ordinary
micro-canonical distribution function ${\tilde Z}_{(nr)}(\mathcal{E}, V, N)$
is equal to the standard one with N-1 particles.

\medskip

For the micro-canonical distribution function we have\medskip

\begin{eqnarray}
f_{(mc,nr)} &&({\vec{\eta}}_{1},..,{\vec{\kappa}}_{N}|\mathcal{E},V,N){|}_{{%
\vec{x}}_{(n)}={\vec{p}}_{(n)}=0}={\tilde{Z}}_{(nr)}^{-1}(\mathcal{E},V,N)\,{%
\frac{{\chi (V)}}{{N!}}}\,\delta (E_{Galilei,N}-\mathcal{E})  \nonumber \\
&&\delta ^{3}({\vec{S}}_{Galilei,N}-{\vec{\mathcal{S}}})\,\delta ^{3}({\vec{%
\mathcal{P}}}_{Galilei,N})\,\delta ^{3}({\frac{{{\vec{\mathcal{K}}}%
_{Galilei,N}}}{{m}}})\approx {\tilde{f}}_{(mc,nr)}({\vec{\rho}}_{(n)1},..,{%
\vec{\pi}}_{(n)N-1}|\mathcal{E},V,N).  \nonumber \\
&&  \nonumber \\
{} &&  \label{5.6}
\end{eqnarray}

\noindent It satisfies the Liouville equation with Hamiltonian $H = \mathcal{%
E} = E_{Galilei}{|}_{{\vec P}_{(n)} = {\vec x}_{(n)} = 0}$ (it corresponds
to the Hamilton-Jacobi description of the center of mass in the rest frame
centered on the center of mass). Moreover it satisfies $\partial_t\, f_{(mc,
nr)} = 0$, so that it is an \textit{equilibrium} distribution function in
statistical mechanics. The statistical average of a function $F(\vec x, \vec
p; {\vec \eta}_i, {\vec \kappa}_i)$ is\medskip

\begin{equation}
F_{(mc, nr)}(\vec x, \vec p | \mathcal{E}, V, N) = \int \prod_{i=1}^N\,
d^3\eta_i\, d^3\kappa_i\, F(\vec x, \vec p; {\vec \eta}_i, {\vec \kappa}%
_i)\, f_{(mc, nr)}({\vec \eta}_1,.., {\vec \kappa}_N |\mathcal{E}, V, N).
\label{5.7}
\end{equation}

\bigskip

In the non-relativistic case, by using the results in Subsection B of
Section II, we can undo the Hamilton-Jacobi transformation on the center of
mass and we can recover Eq.(\ref{5.1}) from the second of Eqs.(\ref{5.5}%
)\medskip

\begin{equation}
{\tilde Z}_{(nr, st)}(E, V, N) = \int d^3x_{(nr)}\, d^3p_{(nr)}\, \theta(R -
|{\vec x}_{(n)}|)\, {\tilde Z}_{(nr)}(E = {\frac{{{\vec p}^2_{(n)}}}{{2m}}}
+ \mathcal{E}, V, N).  \label{5.8}
\end{equation}

\noindent This is possible because the Galilei energy generator is the sum
of the kinetic energy of the center of mass and of the internal energy,
which is an invariant at the non-relativistic level. This property does not
exist at the relativistic level with the Poincar\'e group.

\bigskip

As shown in Appendix C of Ref.\cite{37} (Eqs.(C17) and (C42)-(C43) with ${%
\vec \kappa}_+ = 0$) after a long calculation we get the following
expressions for the standard and the extended micro-canonical distribution
functions in the rest frame for N free particles of mass $m$

\begin{eqnarray}
{\tilde Z}_{(nr)}(\mathcal{E},V,N) &=&\frac{1}{N!}\, \left( \sqrt{\frac{m}{%
2\pi }}\right) ^{3N}\, \frac{\mathcal{E}^{(3N-5)/2}}{\Gamma ((3N-3)/2)}\,
\theta (\mathcal{E})  \nonumber \\
&&\sqrt{\frac{32\pi }{N^{3}m^{9}}} 3^{N-1}V^{N-1}\int_{0}^{\infty
}x^{2}dx\left( \frac{j_{1}(x)}{x}\right) ^{N},  \nonumber \\
&&{}  \label{5.9}
\end{eqnarray}

\begin{eqnarray}
&&{\tilde Z}_{(nr)}( \mathcal{E}, V, N, {\vec {\mathcal{S}}}) =  \nonumber \\
&=&\frac{1}{N!(2\pi )^{9N}}\left( \sqrt{8\pi ^{3}}\right) ^{N+1}\left( \sqrt{
m^{3}}\right) ^{N-1}\frac{(2\pi )^{3}}{\left( m\right) ^{3}}\left( \frac{3V}{
4\pi }\right) ^{N-1}  \nonumber \\
&&\times \prod_{i}^{N}\int_{0}^{1}x_{i}^{2}dx_{i}\int_{0}^{\pi }\sin \theta
_{i}d\theta _{i}\int_{0}^{2\pi }d\phi _{i}\delta ^{3}(\sum_{i=1}^{N}\vec{x}
_{i})\frac{(\mathcal{E} - {\tilde {\mathcal{E}}}({\vec \eta}_i, {\vec {%
\mathcal{S}}}))^{(3N-3)/2-1}} {\Gamma ((3N-3)/2)}\theta (\mathcal{E} - {%
\tilde {\mathcal{E}}}({\vec \eta}_i, {\vec {\mathcal{S}}})),  \nonumber \\
&&{}  \label{5.10}
\end{eqnarray}

\noindent in which

\begin{eqnarray}
{\tilde {\mathcal{E}}}(\vec{\eta}_{i}, {\vec {\mathcal{S}}}) &=&\frac{ (%
\mathcal{S}^1)^2}{2m\sum_{i=1}^{N} (\eta _i^1)^{2}}+\frac{ (\mathcal{S}^2)^2 
}{2m\sum_{i=1}^{N}\, (\eta_i^2)^{2}} + \frac{ (\mathcal{S}^3)^2 }{%
2m\sum_{i=1}^{N}\, (\eta _i^3)^{2}} =  \nonumber \\
&=&\frac{1}{2mR^{2}}\left( \frac{ (\mathcal{S}^1)^2 }{%
\sum_{i=1}^{N}x_{i}^{2}\cos ^{2}\phi _{i}\sin ^{2}\theta _{i}}+\frac{ (%
\mathcal{S}^2)^2 }{\sum_{i=1}^{N}x_{i}^{2} \sin ^{2}\phi _{i}\sin ^{2}\theta
_{i}}+\frac{ (\mathcal{S}^3)^2 }{2m \sum_{i=1}^{N}x_{i}^{2}\cos ^{2}\theta
_{i}}\right).  \nonumber \\
&&{}  \label{5.11}
\end{eqnarray}

\subsection{The Micro-Canonical Ensemble in the Relativistic Inertial Rest
Frame}

Let us remark that in the relativistic case there exists the following
definition of the standard micro-canonical partition function

\begin{equation}
Z_{(st)}(E, V, N) = {\frac{1}{{N!}}}\, \int \chi(V)\, \prod_I^{1..N}\,
d^3x_i\, d^3p_i\, \delta(H_N - E),  \label{5.12}
\end{equation}

\noindent in an arbitrary inertial frame in the free case with $H_N =
\sum_i^{1..N}\, \sqrt{m^2\, c^2 + {\vec p}_i^2}$. Its form is not known in
closed form for $m \not= 0$ (for m=0 see Ref.\cite{27}). This definition is
obtained by describing the N free particles in an inertial frame by means of
their world-lines $x^{\mu}_i$ and of the conjugate momenta $p^{\mu}_i$ by
putting by hand $x^o_1 = ... = x^o_N = x^o$ and by using the mass-shell
conditions $\epsilon\, p^2_i = m_i^2\, c^2$ to eliminate the energies $p^o_i$%
's. This description includes the center of mass but no consistent way to
include interactions among the N particles is known. The inclusion of
interactions was the motivation of the new relativistic classical and
quantum mechanics of Ref.\cite{1} in which the world-lines $x^{\mu}_i$ and
their momenta $p^{\mu}_i$ are derived quantities as shown in Section II. In
the new formulation the Wigner 3-vectors ${\vec \eta}_i(\tau)$ and ${\vec
\kappa}_i(\tau)$ are the fundamental canonical variables together with the
rest-frame conditions.

\bigskip

Instead in this Subsection we define the micro-canonical partition function
(with given internal energy $\mathcal{E}$ and given rest spin ${\vec {%
\mathcal{S}}}$ in the rest frame) inside the instantaneous Wigner 3-spaces
of the rest frame (in the inertial frame centered on the Fokker-Pryce
external 4-center of inertia with 3-velocity $\vec h = 0$) after the
elimination of the internal 3-center of mass. We define everything in the
inertial rest frame but \textit{without including the external center of
mass $\vec z$, $\vec h$}. The new partition function will be defined in
terms of the internal Poincar\'e generators living inside the Wigner
3-spaces $\Sigma_{\tau}$.

\medskip

The natural volume $V$ would be a spherical box centered on the Fokker-Pryce
center of inertia in the Wigner 3-space ($|{\vec \eta}_i(\tau)| \leq R$).
Let $\chi(V) = \prod_i\, \theta(R - |{\vec \eta}_i(\tau)|)$ be the
characteristic function identifying the volume V. However, since the
internal center of mass is eliminated, it is more convenient to use a
characteristic function depending only on the relative variables, namely $%
\chi(V) = \prod_a\, \theta(2R - |{\vec \rho}_a|)$. \medskip

Then the extended and ordinary partition functions of the micro-canonical
ensemble are (in what follows we have $M_N = M_N({\vec \eta}_i, {\vec \kappa}%
_i)$; the Jacobian $J({\vec \rho}_a, {\vec \pi}_a)$ is defined by $\delta^3({%
\frac{{{\vec {\mathcal{K}}}_N}}{{M_N c}}}) = J({\vec \rho}_a, {\vec \pi}%
_a)\, \delta^3(\vec \eta - {\vec \eta}_+({\vec \rho}_a, {\vec \pi}_a)) $)

\begin{eqnarray*}
\tilde Z(\mathcal{E}, {\vec {\mathcal{S}}}, V, N) &=& {\frac{1}{{N!}}}\,
\int\, \prod^{1..N}_i\, d^3\eta_i\, \chi(V)\, \int\, \prod^{1..N}_j\,
d^3\kappa_j\, \delta(M_N\, c^2 - \mathcal{E})  \nonumber \\
&&\delta^3({\vec S}_N - {\vec {\mathcal{S}}})\, \delta^3({\vec {\mathcal{P}}}%
_N)\, \delta^3({\frac{{{\vec {\mathcal{K}}}_N}}{{M_N c}}}) =  \nonumber \\
&&{}  \nonumber \\
&=& {\frac{1}{{N!}}}\, \int\, d^3\eta \prod_{a=1}^{N-1}\, d^3\rho_a\,
\chi(V)\, \int\, \prod_b^{1..N-1}\, d^3\pi_b\,\, J({\vec \rho}_a, {\vec \pi}%
_a)\, \delta^3(\vec \eta - {\vec \eta}_+({\vec \rho}_a, {\vec \pi}_a)) 
\nonumber \\
&&\delta(M_N({\vec \rho}_a, {\vec \pi}_a)\, c^2 - \mathcal{E})\,
\delta^3(\sum_{a=1}^{N-1}\, {\vec \rho}_a \times {\vec \pi}_a - {\vec {%
\mathcal{S}}}),
\end{eqnarray*}

\begin{eqnarray}
\tilde Z(\mathcal{E}, V, N) &=& \int d^3\mathcal{S}\, \tilde Z(\mathcal{E}, {%
\vec {\mathcal{S}}}, V, N) =  \nonumber \\
&=& {\frac{1}{{N!}}}\, \int\, \prod^{1..N}_i\, d^3\eta_i\, \chi(V)\, \int\,
\prod^{1..N}_j\, d^3\kappa_j\, \delta(M_N\, c^2 - \mathcal{E})\, \delta^3({%
\vec {\mathcal{P}}}_N)\, \delta^3({\frac{{{\vec {\mathcal{K}}}_N}}{{M_N c}}}%
) =  \nonumber \\
&&{}  \nonumber \\
&=& {\frac{1}{{N!}}}\, \int\, d^3\eta \prod_{a=1}^{N-1}\, d^3\rho_a\,
\chi(V)\, \int\, \prod_b^{1..N-1}\, d^3\pi_b\, J({\vec \rho}_a, {\vec \pi}%
_a)\, \delta^3(\vec \eta - {\vec \eta}_+({\vec \rho}_a, {\vec \pi}_a)) 
\nonumber \\
&&\delta(M_N({\vec \rho}_a, {\vec \pi}_a)\, c^2 - \mathcal{E}).  \label{5.13}
\end{eqnarray}

Their evaluation should be done with the methods introduced in Appendix C of
Ref.\cite{37} for the non-relativistic case, but the calculations are much
more involved. In particular one should need a closed form for the inverse
Laplace transform of multiple powers of modified Bessel functions.

\bigskip

Let us remark that the 3-vectors ${\vec \eta}_i(\tau)$, ${\vec \kappa}%
_i(\tau)$ are Wigner spin-1 3-vectors so that quantities like ${\vec \kappa}%
_i^2(\tau)$ are Lorentz scalars. The invariant mass $Mc$ (and therefore $%
\mathcal{E}$) is a Lorentz scalar. ${\vec {\mathcal{P}}}$ and ${\vec {%
\mathcal{K}}}/Mc$ are Wigner spin-1 3-vectors: under a Lorentz
transformation $\Lambda$ they undergo a Wigner rotation $R(\Lambda)$, so
that expressions like $\delta^3({\vec {\mathcal{P}}})$ are Lorentz scalars.
Also the rest spin $\vec S$ is a Wigner spin-1 3-vector. Therefore under a
Lorentz transformation we get $\tilde Z(\mathcal{E}, {\vec {\mathcal{S}}},
V, N) \mapsto \tilde Z(\mathcal{E}, R(\Lambda)^{-1}{\vec {\mathcal{S}}}, V,
N)$, ( the volume is a Lorentz scalar because both $|{\vec \eta}_i(\tau)|$
and $|{\vec \rho}_a(\tau)|$ are Lorentz scalars). Instead $\tilde Z(\mathcal{%
E}, V, N)$ is a \textit{Lorentz scalar}. \medskip

\bigskip

Now we have the distribution function

\begin{equation}
f_{(mc)}({\vec \rho}_1,.., {\vec \pi}_{N-1} | \mathcal{E}, V, N) = {\tilde Z}%
^{-1}(\mathcal{E}, V,N)\, {\frac{{\chi(V)}}{{N!}}}\, \delta(M_N\, c^2 - 
\mathcal{E})\, \delta^3({\vec {\mathcal{P}}}_N)\, \delta^3({\frac{{{\vec {%
\mathcal{K}}}_N}}{{M_N c}}}),  \label{5.14}
\end{equation}

\noindent and its extended version

\begin{eqnarray}
{\tilde f}_{(mc)}({\vec \rho}_1,.., {\vec \pi}_{N-1} | \mathcal{E}, {\vec {%
\mathcal{S}}}, V, N) &=& {\tilde Z}^{-1}(\mathcal{E}, {\vec {\mathcal{S}}},
V,N)\, {\frac{{\chi(V)}}{{N!}}}\, \delta(M_N\, c^2 - \mathcal{E})\, \delta^3(%
{\vec S}_N - {\vec {\mathcal{S}}})  \nonumber \\
&& \delta^3({\vec {\mathcal{P}}}_N)\, \delta^3({\frac{{{\vec {\mathcal{K}}}_N%
}}{{M_N c}}}).  \label{5.15}
\end{eqnarray}

\noindent It satisfies the Liouville theorem with $H = Mc$. Moreover it
satisfies $\partial_{\tau}\, f_{(mc)} = 0$, so that it is an \textit{%
equilibrium} distribution function in statistical mechanics. \medskip

Let us now consider a function $F(\vec{\eta},\vec{\kappa};{\vec{\eta}}_{i},{%
\vec{\kappa}}_{i})$ depending from an extra particle present in the Wigner
3-space $\Sigma _{\tau }$ of the rest frame. Its 3-coordinate $\vec{\eta}$
and 3-momentum $\vec{\kappa}$ will depend on the relative variables ${\vec{%
\rho}}_{a}$, ${\vec{\pi}}_{a}$, $a=1,..,N-1$ of the N particles and from
extra canonical relative variables $\vec{\rho}$ and $\vec{\pi}$ so to get a
set of canonical relative variables for N+1 particles. While ${\vec{\eta}}%
_{i}$ and ${\vec{\kappa}}_{i}$ depend only on ${\vec{\rho}}_{a}$, ${\vec{\pi}%
}_{a}$, the variables $\vec{\eta}$ and $\vec{\kappa}$ depend also on $\vec{%
\rho}$ and $\vec{\pi}$. The statistical average of such a function will be a
function only of $\vec{\rho}$ and $\vec{\pi}$

\begin{equation}
\mathcal{F}_{(mc)}(\vec \rho, \vec \pi) = \int \prod_{i=1}^N\, d^3\eta_i\,
d^3\kappa_i\, F(\vec \eta, \vec \kappa; {\vec \eta}_i, {\vec \kappa}_i)\,
f_{(mc)}({\vec \rho}_1,.., {\vec \pi}_{N-1} | \mathcal{E}, V, N).
\label{5.16}
\end{equation}

\noindent describing the statistical behavior of the relative variables $%
\vec \rho$ and $\vec \pi$ in the Wigner rest 3-space under the influence of
the other N-1 pairs.

\subsection{The Micro-Canonical Ensemble in the Relativistic Inertial Rest
Frame as seen by an Arbitrary Inertial Observer}

Unlike with the non-relativistic case we cannot reintroduce the external
center of mass by using the frozen Jacobi data $\vec z$, $\vec h$: even if
the measure $d^3z\, d^3h$ is Lorentz invariant, $\vec z$ is a non-covariant
quantity.

\medskip

The results at the end of Subsection A of Section II imply that only \textit{%
on-shell} one could rewrite the distribution function $f_{(mc)}({\vec \rho}%
_1,.., {\vec \pi}_{N-1} | \mathcal{E}, V, N)$ as a function of the particle
world-lines $x_i^{\mu}(\tau)$ and of their momenta $p_i^{\mu}(\tau)$.
Moreover in each inertial frame the new distribution function would depend
on the non-covariant pseudo-world-line of the external center of mass, i.e.
on the non-covariant Jacobi data $\vec z$. We cannot put $\vec z = 0$ in a
frame-independent way like in the non-relativistic case. There is a spatial
non-separability implied by the nature of the relativistic collective
variable (due to the Lorentz signature of the space-time) and by the
elimination of relative times in relativistic bound states. One has to learn
to think only in terms of relative 3-variables.

\medskip

Therefore a manifestly covariant micro-canonical distribution function of
the type $F_{(mc)}({\vec {\tilde x}}_i(x^o), {\vec {\tilde p}}_i(x^o) | 
\mathcal{E}, V, N)$, i.e. depending on the world-lines and their momenta in
an arbitrary Lorentz frame (like in all the existing approaches) does not
exist. In group-theoretical terms the basic obstruction to get this type of
distribution function is that the Poincar\'e energy cannot be written as the
center-of-mass energy plus an internal energy like in the case of the
Galilei group.

\subsection{The Micro-Canonical Temperature in the Non-Relativistic and
Relativistic Inertial Rest-Frame}

In the standard non-relativistic micro-canonical ensemble the \textit{%
micro-canonical entropy} is\medskip

\begin{equation}
S_{(mc,nr,st)}(E,V,N)=\,ln\,{\tilde{Z}}_{(nr,st)}(E,V,N).  \label{5.20}
\end{equation}

\medskip

\noindent and the \textit{micro-canonical temperature} $%
T_{(mc)}=T_{(mc)}(E,V,N)$ (see Refs.\cite{27,28,29}; $k_{B}$ is the
Boltzmann constant) is\medskip\ 

\begin{equation}
{\frac{1}{{k_B\, T_{(mc)}}}} = {\frac{{\partial\, S_{(mc,nr,st)}(E,V,N)}}{{%
\partial\, E}}}{|}_{V,N} = {\frac{1}{{{\tilde Z}_{(nr,st)}(E,V, N)}}}\, {%
\frac{{\partial\, {\tilde Z}_{(nr,st)}(E,V,N)}}{{\partial\, E}}}{|}_{V, N}.
\label{5.21}
\end{equation}

\bigskip

Then one can introduce the Gibb's relation $dE = T_{(mc)}\, dS_{(mc,nr,st)}
- P_{(mc)}\, dV + \mu_{(mc)}\, dN$ ($\mu_{(mc)}$ chemical potential) with ${%
\frac{{P_{(mc)}}}{{k_B\, T_{(mc)}}}} = {\frac{{\partial\, S_{(mc)}(E,V,N)}}{{%
\partial\, V}}}{|}_{E,N}$ and ${\frac{{\mu_{(mc)}}}{{k_B\, T_{(mc)}}}} = {%
\frac{{\partial\, S_{(mc)}(E,V,N)}}{{\partial\, N}}}{|}_{E,V}$, and the
second law of thermodynamics $dS_{(mc,nr,st)} \geq 0$. With long range
forces the micro-canonical ensemble is inequivalent to the canonical
ensemble (in which there is negative heat capacity ) as shown Refs. \cite{25}%
.

\bigskip

These definitions can be adapted to the non-relativistic rest frame and then
extended to the relativistic rest frame by replacing the micro-canonical
entropy (\ref{5.20}) with the entropies $S_{(mc,nr)}(\mathcal{E},V,N)=\,ln\,{%
\tilde{Z}}_{(nr)}(\mathcal{E},V,N)$ and $S_{(mc)}(\mathcal{E},V,N)=\,ln\,%
\tilde{Z}(\mathcal{E},V,N)$, respectively. \medskip 

This implies that in the relativistic inertial rest frame the
micro-canonical temperature $T_{(mc)}$ (${\frac{1}{{k_B\, T_{(mc)}}}} = {%
\frac{1}{{\tilde Z(\mathcal{E},V, N)}}}\, {\frac{{\partial\, \tilde Z(%
\mathcal{E},V,N)}}{{\partial\, \mathcal{E}}}}{|}_{V, N}$) is a \textit{%
Lorentz scalar}, because the relativistic internal energy $\mathcal{E}$ is a
Lorentz scalar like the internal energy $M c^2$. \medskip

Therefore in the short range case (equivalence of the micro-canonical and
canonical ensembles) the thermodynamic limit $N, V\,\, \rightarrow\,\,
\infty $ with $N/V = const.$ gives rise to a canonical temperature $T$,
limit of $T_{(mc)}(E, V, N)$, which is a \textit{Lorentz scalar}. Therefore
in the relativistic rest-frame instant form of dynamics we have $T =
T_{rest} $ (see the Introduction for the existing three points of view).

\bigskip

In the case of the ideal Boltzmann gas (N free non-relativistic particles of
mass m and energy ${\frac{{{\vec p}^2}}{{2m}}}$) Eqs. (\ref{5.21}) and (\ref%
{5.3}) imply ${\frac{1}{{k_B\, T_{(mc)}}}}\, {\rightarrow}_{N \rightarrow
\infty}\, {\frac{{3 N}}{{2 E}}}$ (like in the classical virial theorem).
Moreover one gets $p_{(mc)} = k_B\, T_{(mc)}\, {\frac{{\partial\,
S_{(mc,nr,st)}(E, V, N) }}{{\partial\, V}}}{|}_{E,N} {\rightarrow}_{N
\rightarrow \infty}\, k_B\, T_{(mc)}\, {\frac{N}{V}}$ and the resulting
equation of state is $p_{(mc)}\, V = N\, k_B\, T_{(mc)}$. When the
thermodynamics limit is well defined, then $T_{(mc)}$ and $p_{(mc)}$ become
the canonical temperature and pressure, respectively.

\medskip

These results can be reproduced also in the non-relativistic inertial rest
frame by replacing Eq.(\ref{5.3}) with Eq.(\ref{5.9}) as shown in Subsection
5 of Appendix C of Ref.\cite{37}. \medskip

However, in the relativistic rest frame we are not able to find an explicit
analytic form of the micro-canonical entropy $S_{(mc)}(\mathcal{E}%
,V,N)=\,ln\,\tilde{Z}(\mathcal{E},V,N)$ (see Subsection 6 of Appendix C of
Ref.\cite{37}), so that we cannot obtain an explicit equation of state for a
relativistic ideal Boltzmann gas (N free relativistic particles of mass $m$
and energy $\sqrt{m^{2}c^{2}+{\vec{\kappa}}^{2}}$). However in Ref.\cite{14}%
, by using the equilibrium J\"{u}ttner one particle distribution function,
it is shown that also in the relativistic case one obtains $p={\frac{N}{V}}%
\,k_{B}\,T$.

\subsection{The Micro-Canonical Ensemble in the Relativistic Non-Inertial
Rest Frame}

Let us extend the definition (\ref{5.13}) of the extended micro-canonical
partition function to the relativistic non-inertial rest frames by using the
10 asymptotic Poincar\'e generators given in Eqs.(\ref{3.9}) and (\ref{3.10}%
) (see also Ref.\cite{3}) and the non-inertial Hamiltonian $\mathcal{M} c$
of Eq.(\ref{3.11}). The non-inertial extended partition function, depending
on the inertial potentials $g$ and $g^r$ appearing in Eq.(\ref{3.7}), is

\begin{eqnarray}
\tilde Z(\mathcal{E}, {\vec {\mathcal{S}}}, V, N | g, g^r) &=& {\frac{1}{{N!}%
}}\, \int \chi(V)\, \prod_{i=1}^{N}\, d^3\eta_i\, d^3\kappa_i\,\, \delta(%
\mathcal{M}_N({\vec \eta}_i, {\vec \kappa}_i; g, g^r)\, c^2 - \mathcal{E}) 
\nonumber \\
&& \delta^3(J^u_N({\vec \eta}_i, {\vec \kappa}_i; g, g^r) - \mathcal{S}^u)
\, \delta^3({\hat {\mathcal{P}}}_N^u({\vec \eta}_i, {\vec \kappa}_i; g,
g^r))\, \delta^3({\frac{{{\hat {\mathcal{K}}}_N^u({\vec \eta}_i, {\vec \kappa%
}_i; g, g^r)}}{{\mathcal{M}_N({\vec \eta}_i, {\vec \kappa}_i; g, g^r) c}}}),
\nonumber \\
&&{}  \label{5.22}
\end{eqnarray}

\noindent

Let us remark that now we cannot introduce relative variables, because they
are not tensorially defined in non-Euclidean 3-spaces. In the interacting
case with potentials $V({\vec \eta}_i(\tau) - {\vec \eta}_j(\tau); {\vec
\kappa}_i(\tau)) = \tilde V(\sqrt{({\vec \eta}_i(\tau) - {\vec \eta}%
_j(\tau))^2}; {\vec \kappa}_i(\tau))$ we must replace the quantity $({\vec
\eta}_i(\tau) - {\vec \eta}_j(\tau))^2$ with the bi-scalar Synge world
function for Riemannian 3-spaces $\sigma_{(ij)}({\vec \eta}_i(\tau), {\vec
\eta}_j(\tau)) = {\frac{1}{2}}\, (\lambda_1 - \lambda_0)\,
\int_{\lambda_o}^{\lambda_1}\, d\lambda\,\, {}^3g_{rs}(\eta_{(ij)}(\lambda,
\tau))\, {\frac{{\partial \eta^r_{(ij)}(\lambda, \tau}}{{\partial\, \lambda}}%
}\, {\frac{{\partial \eta^s_{(ij)}(\lambda, \tau}}{{\partial\, \lambda}}}$,
where $\eta_{(ij)}(\lambda, \tau)$ is the 3-geodesic joining ${\vec \eta}%
_i(\tau)$ and ${\vec \eta}_j(\tau)$ (see Ref. \cite{27}). The momenta are
covectors defined at the positions of the particles \bigskip

\bigskip

In this case the distribution function is

\begin{eqnarray}
&&{\tilde f}_{(mc)}({\vec \rho}_1,.., {\vec \pi}_{N-1}| \mathcal{E}, {\vec {%
\mathcal{S}}}, V, N | g, g^r) = {\tilde Z}^{-1}(\mathcal{E}, {\vec {\mathcal{%
S}}}, V,N | g, g^r)\, \chi(V)\, \delta(M_N({\vec \eta}_i, {\vec \kappa}_i;
g, g^r)\, c^2 - \mathcal{E})  \nonumber \\
&&{}  \nonumber \\
&& \delta^3(J^u_N({\vec \eta}_i, {\vec \kappa}_i; g, g^r) - \mathcal{S}^u)
\, \delta^3({\hat {\mathcal{P}}}_N^u({\vec \eta}_i, {\vec \kappa}_i; g,
g^r))\, \delta^3({\frac{{{\hat {\mathcal{K}}}_N^u({\vec \eta}_i, {\vec \kappa%
}_i; g, g^r)}}{{M_N({\vec \eta}_i, {\vec \kappa}_i; g, g^r) c}}}).  \nonumber
\\
&&{}  \label{5.23}
\end{eqnarray}

\medskip

It satisfies the Liouville theorem with the Hamiltonian $\mathcal{M}$ of Eq.(%
\ref{3.11}). In the non-inertial rest frames $J^u_N({\vec \eta}_i, {\vec
\kappa}_i; g, g^r)$, ${\hat {\mathcal{P}}}_N^u({\vec \eta}_i, {\vec \kappa}%
_i; g, g^r)$, ${\frac{{{\hat {\mathcal{K}}}_N^u({\vec \eta}_i, {\vec \kappa}%
_i; g, g^r)}}{{M_N({\vec \eta}_i, {\vec \kappa}_i; g, g^r) c}}}$ are
asymptotic constants of the motion at spatial infinity. As a consequence we
have $\partial_{\tau}\, f_{(mc)} = 0$ notwithstanding the presence of the
time-dependent long-range inertial potentials. Therefore in this passive
viewpoint (we do not actively accelerate the gas but we go passively from an
inertial to a non-inertial frame) we get an equilibrium distribution
function also in non-inertial rest frames in accord with their gauge
equivalence to the inertial ones shown in Section II.

\medskip

By using a definition of entropy like in Eq.(\ref{5.20}) the micro-canonical
temperature turns out to be a functional of the inertial potentials.

\subsection{The Micro-Canonical Ensemble in the Non-Relativistic
Non-Inertial Rest Frame}

By using the results of Subsection D of Section II for the non-inertial
Galilei generators we get the following definition of the non-relativistic
extended partition function in non-inertial rest frames

\begin{eqnarray}
{\tilde Z}_{(nr)}(\mathcal{E}, {\vec {\mathcal{S}}}, V | g, g^r) &=& {\frac{1%
}{{N!}}}\, \int\, \prod^{1..N}_i\, d^3\eta_i\, \chi(V)\, \int\,
\prod^{1..N}_j\, d^3\kappa_j\, \delta(E_{Galilei, N} - \mathcal{E}) 
\nonumber \\
&&\delta^3({\vec S}_{Galilei, N} - {\vec {\mathcal{S}}})\, \delta^3({\vec {%
\mathcal{P}}}_{Galilei, N})\, \delta^3({\frac{{{\vec {\mathcal{K}}}%
_{Galilei, N}}}{{m}}}).  \label{5.24}
\end{eqnarray}

\bigskip

For the distribution function we have

\begin{eqnarray}
{\tilde f}_{(mc, nr)}({\vec \rho}_1,.., {\vec \pi}_{N-1} | \mathcal{E}, {%
\vec {\mathcal{S}}}, V, N | g, g^r) &=& {\tilde Z}_{(nr)}^{-1}(\mathcal{E}, {%
\vec {\mathcal{S}}}, V,N | g, g^r)\, \chi(V)\, \delta(E_{Galilei, N} - 
\mathcal{E})  \nonumber \\
&& \delta^3({\vec S}_{Galilei, N} - {\vec {\mathcal{S}}})\, \delta^3({\vec {%
\mathcal{P}}}_{Galilei, N})\, \delta^3({\frac{{{\vec {\mathcal{K}}}%
_{Galilei, N}}}{{m}}}).  \nonumber \\
&&{}  \label{5.25}
\end{eqnarray}

\medskip

By using the Hamilton equations generated by the Hamiltonian (\ref{3.17}) it
can be checked that it satisfies the Liouville theorem. Due to the
asymptotic constancy of the Galilei generators at spatial infinity we get
again an equilibrium distribution function, $\partial_t\, {\tilde f}_{(mc,
nr)} = 0$, like in inertial frames.

\section{Conclusions}

Our new consistent RCM allows us to revisit long standing problems in
relativistic statistical mechanics and relativistic kinetic theory. \medskip

The first part of the paper is devoted to a clarification of the kinematical
background which was the basic tool to develop a relativistic theory of
isolated systems in accord with what is known about relativistic bound
states and using a methodology for the definition of the instantaneous
3-spaces which is a mathematical idealization of the protocols for clock
synchronization used in atomic and space physics. In this way we can arrive
at a description of N-particle systems with various kinds of interactions
and with an explicit knowledge of the Poincar\'{e} generators and a control
on the relativistic collective variables in Minkowski space-time. The
non-relativistic limit allows us to recover a Hamilton-Jacobi description of
the Galilei generators in Galilei space-time. \medskip

After a study of the inertial rest frames of isolated systems we also gave
their description in the non-inertial rest frames, both at the relativistic
and non-relativistic level. \medskip

The main effect of the elimination of relative times in relativistic bound
states and of the decoupling of the external non-covariant (non-local and
therefore non-measurable) center of mass is the introduction of a \textit{%
spatial non-separability} forcing us to describe every isolated system only
in terms of Wigner-covariant relative 3-variables inside the 3-spaces. A
manifestly-covariant description of the distribution functions of
relativistic statistical mechanics in terms of the world-lines of the
particles and their 4-momenta is not possible. Besides this spatial
non-separability there is the non-measurability of the decoupled external
non-covariant canonical center of mass due to its non-local nature. See Ref.%
\cite{41} for more details on this point and for the indications from
non-relativistic quantum mechanics that also the localization of the Newton
center of mass is an open problem.

\medskip

The main result of the paper is the use of this framework to give a
definition of the micro-canonical ensemble both in relativistic and
non-relativistic statistical mechanics which utilizes only the Poincar\'e or
Galilei generators. The definition is given in the inertial rest frames, but
it can be extended to the non-inertial ones (till now no such an extension
was known) in particular to the non-inertial rest frames. The so defined
non-inertial micro-canonical ensemble turns out to describe an equilibrium
configuration notwithstanding the presence of long-range inertial potentials
due to the fact the asymptotic Poincar\'e or Galilei generators are
constants of the motion in non-inertial rest frames. \medskip

With our definition of the relativistic micro-canonical ensemble the
micro-canonical temperature turns out to be a Lorentz scalar. Since we do
not discriminate between long- and short-range interactions, we have no
statement about the relativistic canonical ensemble, except that when the
thermodynamic limit exists also the canonical temperature is a Lorentz
scalar. We have no new statement about the definition of relativistic
thermodynamics.\medskip

\bigskip

Let us finish by adding that in the extended version of Ref.\cite{37} there
is also a preliminary investigation of the following open
problems:\hfill\break

A) Starting from the density function of a non-equilibrium relativistic
Gibbs ensemble (assumed to transform like the relativistic micro-canonical
distribution function to which it tends at equilibrium) we can give a
definition of the one-particle distribution function $f(\vec \eta, \vec
\kappa, \tau)$ with a statistical average. We only investigate the
transformation properties of this one-particle distribution function under
Poincar\'e transformations in the inertial relativistic rest frame. We can
show that it is a \textit{Lorentz scalar}: this is our answer to the long
standing debate on the transformation properties of this distribution
reviewed in Refs. \cite{42,43,44}.\medskip

In the inertial non-relativistic case one can recover the Maxwell-Boltzmann
distribution function as the equilibrium solution of the non-relativistic
Boltzmann equation, which can be derived (see for instance chapter 3 of Ref.%
\cite{13}) as an approximation from the BBGKY (Bogoliubov - Born - Green -
Kirkwood - Yvon) hierarchy for the coupled equations of motion of the
s-particle distributions functions implied by the Liouville theorem using
the Hamiltonian of the N-particle system. In the relativistic case in
absence of a consistent RCM the Boltzmann equation is either postulated or
derived from quantum field theory (see Refs.\cite{45}, \cite{14}, \cite{46}) 
\footnote{%
Starting from the existing relativistic kinetic theory \cite{47,48} one
arrives at the \textit{relativistic Boltzmann equation} of Ref.\cite{46}. As
shown in Ref.\cite{17} if we consider a gas of charged massive particles
interacting with external electro-magnetic and gravitational fields, the
Hamilton equations of motion of a particle are ${\frac{{d x^{\mu}(\lambda)}}{%
{d \lambda}}} = p^{\mu}(\lambda)$, ${\frac{{d p^{\mu}(\lambda)}}{{d \lambda}}%
} = e\, F^{\mu}{}_{\nu}(x(\lambda))\, p^{\nu}(\lambda) -
\Gamma^{\mu}_{\alpha\beta}(x(\lambda))\, p^{\alpha}(\lambda)\,
p^{\beta}(\lambda) = F^{\mu}(x(\lambda), p(\lambda))$ ($\lambda$ is an
affine parameter; self-forces are not considered). Therefore, if $\hat L =
p^{\mu}\, {\frac{{\partial}}{{\partial\, x^{\mu}}}} + F^{\mu}\, {\frac{{%
\partial}}{{\partial\, p^{\mu}}}}$ is the associated Liouville operator, the
relativistic Boltzmann equation is ${\hat L}{|}_{\epsilon\,
g^{\alpha\beta}(x)\, p_{\alpha}\, p_{\beta} = m^2\, c^2}\, f(x^o, \vec x,
\vec p) = \mathcal{C}[f]$, where $\mathcal{C}[f]$ is the collision term,
bilinear in $f$ for 2-body collisions. At equilibrium (no collisions) the
distribution function $f_{eq}(x^o, \vec x, \vec p)$ satisfies this equation
with $\mathcal{C}[f_{eq}] = 0$ and $\partial_{x^o}\, f_{eq} = 0$.}, and for
free particles the equilibrium solution is the relativistic
Boltzmann-J\"uttner distribution function \cite{49}. By adapting this
construction to the relativistic inertial rest frame, we show that the
J\"uttner distribution can be obtained also in our approach.

\medskip

Since in our RCM we have the Hamilton equations for the interacting
N-particle systems of Section IV (and of Appendices A and B) of Ref.\cite{37}
described by Wigner-covariant phase space 3-variables ${\vec \eta}_i(\tau)$, 
${\vec \kappa}_i(\tau)$, $i=1,..,N$, (or by the relative ones ${\vec \rho}%
_a(\tau)$, ${\vec \pi}_a(\tau)$, $a=1,..,N-1$, after the elimination of the
internal center of mass in the Wigner 3-spaces), we can introduce also the
correlation functions $f_s({\vec \eta}_1, {\vec \kappa}_1,..,{\vec \eta}_s, {%
\vec \kappa}_s, \tau)$, $s=1,..,N$, and, by using the Liouville operator
associated to the relativistic N-particle Hamiltonian (the global invariant
mass), find their coupled equations of motion. This produces a relativistic
BBGKY hierarchy, from which the form of the relativistic Boltzmann equation
implied by our RCM, with a decoupled external relativistic center of mass,
could be defined in absence of external forces. However the problem is much
more complex than in the non-relativistic case due to the form of the
relativistic Hamiltonian and to the \textit{momentum-dependence of the
potentials} and only for certain models and with drastic approximations can
a Boltzmann equation be derived.

\bigskip

B) The moments of the one-particle distribution function, solution of the
Boltzmann equation, give rise to a hydrodynamical description of
relativistic kinetic theory with an effective dissipative fluid (to be
contrasted with the perfect or dissipative fluids of relativistic
hydrodynamics). This allows us to study the problems of relativistic
dissipative fluids and of causal relativistic thermodynamics, whose
foundations are not yet fully established \footnote{%
For instance there are discussions going on whether the standard definition
of relativistic thermodynamics given in Ref.\cite{47}) is acceptable. See
for instance Ref.\cite{50} for a proposal to modify the first member of the
Gibbs relation $de = Tds + \mu\, dn$ by adding a dependence on the momentum
density to the variation $de$ of the local rest frame energy density.}, but
which is used for instance as a hydrodynamical model for describing
relativistic heavy-ion collisions \cite{51}. See Refs.\cite{52,53} for a
review of the Eckart, Landau-Lifschitz, Israel-Stewart \cite{47} and Carter 
\cite{54} points of view and Ref.\cite{55} for the 1+3 point of view 
\footnote{%
In this paper we use a 3+1 point of view, which is defined in Subsection A
of Section II. Instead in the 1+3 point of view one gives only the
world-line of a time-like observer and the instantaneous 3-spaces are
identified with the tangent spaces orthogonal to the unit 4-velocity of the
observer in each point of the world-line. As shown in Ref.\cite{3} these
tangent planes intersect each other at a certain distance from the observer,
so that coordinates like the Fermi or rotating ones can be defined only
locally due to these coordinate singularities.}.\medskip

The various approaches differ in the parametrization of the relativistic
entropy current $S^{\mu} = s\, u^{\mu} + q^{\mu} + V^{\mu}$ in terms of the
fluid 4-velocity $u^{\mu}$, the ordinary entropy $s$, the heat transfer
4-vector $q^{\mu}$ and the viscous terms $V^{\mu}$ (with $q^{\mu}$ and $%
V^{\mu}$ orthogonal to $u^{\mu}$), in the definition of $u^{\mu}$ and in the
order of the deviations from equilibrium (linear for the Eckart and
Landau-Lifschitz models, quadratic for the Israel-Stewart and Carter
models). See Refs.\cite{48} and the Appendix B of Ref.\cite{56} for reviews;
for recent developments see Refs.\cite{57,58,59}.

Then one has to implement the second law of thermodynamics in the form $%
\partial_{\mu}\, S^{\mu} \geq 0$. In Carter's two-fluid approach one writes $%
S^{\mu} = s\, u^{\mu} + j^{\mu}$ and considers $j^{\mu}$ as a second fluid;
a Lagrangian-like approach is used to implement $\partial_{\mu}\, S^{\mu}
\geq 0$. As shown in Ref.\cite{60} Israel-Stewart and Carter approaches are
essentially equivalent.

What is lacking in all these approaches is a variational principle
describing relativistic fluids out of equilibrium and implying $%
\partial_{\mu}\, S^{\mu} \geq 0$. Our last contribution to the relativistic
framework is the suggestion that the action principle for relativistic
fluids of Ref.\cite{61}, in the version of Refs.\cite{56,62,63}, can be
modified to get these results.

\vfill\eject

\end{document}